\documentclass{eas}
\usepackage{graphicx}
\usepackage{natbib}
\usepackage{color}
\usepackage{amsmath}
\usepackage{url}
\usepackage{enumitem}
\newcommand\bb[1] {   \mbox{\boldmath{$#1$}}  }
\newcommand\del{\bb{\nabla}}
\newcommand\bcdot{\bb{\cdot}}
\newcommand\btimes{\bb{\times}}
\newcommand{\mean}[1]{\langle #1 \rangle}

\begin{document}

\title{MRI--driven angular momentum transport in protoplanetary disks} 
\runningtitle{MRI--driven MHD turbulence in accretion disks}
\author{S\'ebastien Fromang}\address{Laboratoire AIM,
  CEA/DSM-CNRS-Universit\'e Paris Diderot, IRFU/Service
  d'Astrophysique, CEA--Saclay F-91191 Gif-sur-Yvette, France}
\begin{abstract}
Angular momentum transport in accretion disk has been the focus of
intense research in theoretical astrophysics for many decades. In the
past twenty years, MHD turbulence driven by the magnetorotational
instability has emerged as an efficient mechanism to achieve that
goal. Yet, many questions and uncertainties remain, among which the
saturation level of the turbulence. The consequences of the
magnetorotational instability for planet formation models are still
being investigated. This lecture, given in September 2012 at the school
"Role and mechanisms of angular momentum transport in the formation
and early evolution of stars'' in Aussois (France), aims at
introducing the historical developments, current status and outstanding
questions related to the magnetorotational instability that are
currently at the forefront of academic research.
\end{abstract}
\maketitle
\section{Introduction}
\label{intro_sec}

\begin{quotation}
{\it ``In this paper and a companion work, we show that a broad class of
astrophysical accretion disk is dynamically unstable to axisymmetric
disturbances in the presence of a weak magnetic field.''}
\citep{balbus&hawley91} 
\end{quotation}

\noindent
The above sentence is opening the soon to be seminal paper of
\citet{balbus&hawley91}. It brought to the scene a magnetohydrodynamical
instability that destabilizes magnetized accretion disks. Although
known since the 60's \citep{velikhov59,chandrabookhydro}, its
importance in the field of astrophysics had not been appreciated. In
the few years that followed, it was rapidly realized that the
magnetorotational instability (or, in short, MRI), as it was to be
called, has profound consequences for the dynamics of accretion 
disks. It is now believed to be the main physical effect responsible
for the fall of matter onto the object (young star, white dwarf,
neutron star or black hole) that sits at the center of accretion
disks, thus solving a long lasting mystery of modern astrophysics. Its
consequences cannot be ignored when studying such important issues as
planet formation or black hole physics. In just two decades, the
paper by \citet{balbus&hawley91} changed the field of accretion disk
astrophysics.

\noindent
This lecture is aimed at PhD students, postdocs and researchers that
are non--specialist of that field. Its purpose is to establish as
simply as possible the basic physical mechanisms associated with the
MRI come into play and to highlight the outstanding questions we
currently struggle with. Even if some aspects of the lecture have
broad implications to all categories of accretion disks, it tends 
to be focused on the particular case of protoplanetary (PP)
disks, i.e. accretion disks rotating around newly born stars. Since PP
disks are the nurseries of planetary systems such as our own, research
devoted to understanding their properties has been intense ever since 
the discovery of the MRI. As we shall see, the conditions that prevail
in PP disks result in peculiarities that greatly complicate their
dynamics. Many outstanding questions remain for future generations to
solve. 

Before moving on, let me emphasize that this lecture is not a review
about the MRI. It is meant to be an introduction to the topic at the
most basic level and it is strongly biased toward PP disks. For a more
complete overview of the MRI, especially for those issues that are
specific to other classes of disks than just PP disks, 
the interested reader will benefit from the published reviews by
\citet{balbus&hawley98,balbus&hawley00,balbusaraa03}. The subject of
PP disks itself is very rich and many aspects of PP disks not related
to the MRI are left aside in this lecture. Excellent reviews 
on PP disks have recently been published that provide a
comprehensive overview of their properties. The interested reader is
referred for example to \citet{dullemond&monnier10} or 
\citet{armitage11} for completeness.

\paragraph{Angular momentum transport in accretion disks} Before
moving to the physics of the MRI and its consequences, it is first
important to understand the stakes of the problem. What are we after?
The question we are facing is quite simple: matter in accretion
disks falls onto the central object, and we want to understand
why. The answer is not as simple as it may sound, though! To first
order, matter in accretion disks describes circular orbits around the
central object (we will note its mass $M$ in the following). This is
much like the earth orbiting around the sun. The 
force balance is between the central gravitational attraction and the
centrifugal force. In Newtonian dynamics, that balance results in the
orbital velocity $v_\phi$ of each element being Keplerian:
\begin{equation}
v_\phi=R\Omega=\sqrt{\frac{GM}{R}} \, .
\end{equation}
In the above equation, $G$ is the gravitational constant and $R$ is
the distance to the central object. The parameter $\Omega$ is called
the angular velocity and is an important parameter of this course. The 
important point of that formula, though, is that matter possesses a
finite amount of angular momentum per unit mass,
$L=Rv_\phi=R^2\Omega$, that it must 
lose in order to be accreted onto the central object. This requires
dissipative effects that are able to extract angular momentum from
each element. Otherwise, angular momentum is conserved and matter,
like the earth, will stay for ever on its circular Keplerian
orbit. The problem of mass accretion, thus, translates into the
central theme of this book: angular momentum transport.

\noindent
How can angular momentum be extracted from matter in accretion disks?
The answer to that question is easy, isn't it? As we just saw,
accretion disks are in Keplerian rotation, which means the angular
velocity is varying with  
radius. Or, in other words, the flow is sheared: elements sitting on
neighboring orbits rotate at different rates. This is fortunate as
we know from fluid mechanics courses that the fluid molecular
viscosity $\nu$ exerts a force between two differentially rotating
rings. The inner ring is slowed down by the outer (and more slowly
rotating) ring. Stated in terms of angular momentum conservation,
this means that angular momentum flows from the inner ring to the outer
ring: the inner ring's angular momentum decreases and the ring itself
moves in as a result. In other words it accretes toward the central
object. The problem of accretion disk is solved! Unfortunately, things
are not that simple and it turns out that this naive picture badly
fails. Consider indeed the timescale associated with this diffusive
process. The typical time to transport matter over the distance $R$ is
given by the viscous timescale
\begin{equation}
t_\nu=\frac{R^2}{\nu} \, .
\end{equation}
An order of magnitude estimate of the viscosity can be obtained as
follows. It is of the order of the thermal velocity of the molecules
$v_{th}$ times their mean free path $\lambda$: $\nu \sim v_{th}
\lambda$. $v_{th}$ can be estimated, for molecules of mass $m$ in a
medium at temperature T, through the relation 
\begin{equation}
\frac{1}{2}m_{H_2}v_{th}^2 \sim \frac{3}{2} k_BT \, ,
\end{equation}
where $k_B$ is the Boltzmann constant. For hydrogen molecules (of mass
$m_{H_2}=3.4 \times 10^{-27}$ kg) at a temperature of about $100$ K
typical of PP disks, we find $v_{th} \sim 1$ km.s$^{-1}$. The
molecules mean free path can be approximated by
\begin{equation}
\lambda \sim \frac{1}{n_{H_2}\sigma} \, .
\end{equation}
Here, $\sigma \sim \pi a^2 \sim 3 \times 10^{-16}$ cm$^{-2}$ is the
geometrical cross section between molecules (we have taken $a=1$
Angstr\"om as a typical molecular size). $n_{H_2}$ is the number of
molecules per unit volume. If we 
call $\Sigma$ the disk surface density and $H$ its thickness, then
$n_{H_2}m_{H_2} \sim \Sigma/H$. In PP disks, typical values of the
surface density at $1$ Astronomical Unit (AU) from the central star
are of order $10^3$ g.cm$^{-2}$, while $H/R \sim 0.05$. Thus we find, at
$1$ AU,  $n_{H_2} \sim 4 \times 10^{14}$ cm$^{-3}$. Putting things
together, we obtain $\lambda \sim 0.1$--$1$ m\footnote{We note in
passing that the mean free path of the gas is much smaller than the
dimension of the system. We will thus adopt a fluid description to
described its dynamics at the scales of interest (typically of order a
fraction of $H$) for this lecture.}, from which $\nu \sim 10^3$
m$^2$.s$^{-1}$ follows. We are 
now in a position to evaluate the viscous timescale of PP disks. At
$1$ AU, we obtain $t_\nu \sim 10^{18}$ s. This is about $10^{11}$ yr,
i.e. much longer than the age of the universe! Surely, a more
efficient mechanism must exist to account for the observation that PP
disks lifetime amounts to a few million years.

\noindent
Our intuition based on laboratory experiment as well as everyday life
is that turbulence could well be such a mechanism. Indeed, turbulence is
known to be efficient at transporting things around. In the same way the
turbulent air in a room can rapidly transport molecules of perfume from
one side of that room to the other, turbulence in accretion disks
could well transport angular momentum from their inner regions to
their outer parts. This is why people quickly made the ansatz that
accretion disks are turbulent. This hypothesis is further supported by
the realization that the Reynolds number $Re$ of the flow is gigantic
in accretion disks. Using the above estimates for the various physical
parameters of the system, $Re$ can indeed be estimated to be
\begin{equation}
Re=\frac{v_\phi H}{\nu} \sim 10^{11} \gg 1\, .
\end{equation}
Laboratory experiments tell us that fluid flows become turbulent at high
Reynolds number, even when they are linearly stable as is the case of
Keplerian flow (a result known as the Rayleigh criterion that we will
recover in section~\ref{mri_disp_sec}). This is known as nonlinear
instability. Is it also the case of accretion disks? Well, despite decades
of intense research and vivid controversy, the nonlinear stability of
accretion disks is still debated and there is no accepted argument, be
it theoretical, numerical or experimental, that accretion disks are
hydrodynamically turbulent. We shall return to that issue in more
details in section~\ref{hydro_transport_sec}. For the purpose of this
introduction, it is sufficient to note that the source of 
turbulence in accretion disks remained elusive for many years. 

\noindent
Models of accretion disks, however, were made possible by the
introduction in the 70's of the so--called $\alpha$ disks model in 
two seminal papers written by \citet{shakura&sunyaev73} and
\citet{LP74}. Such models are in fact large scale models of the
turbulence, different versions of which are commonly used in other 
field of physics. The idea is to replace the molecular viscosity
discussed above by an anomalous (i.e. boosted) viscosity that is
supposed to mimic the effect of turbulence. How unsatisfactory as it
may seem, such an approach generally give helpful insights,
qualitative and even sometimes quantitative agreements with real
flows. As we saw above, viscosity has the dimensions of a 
velocity times a length. Using the sound speed $c_s$ as a characteristic
velocity and the disk thickness as a characteristic length,
\citet{shakura&sunyaev73} and \citet{LP74} introduced the following
scaling for the viscosity 
\begin{equation}
\nu_t=\alpha c_s H \, ,
\label{alpha_def_eq}
\end{equation}
where $\alpha$ is a dimensionless quantity, supposedly constant, after
which the model is named. The subscript $t$ serves to differentiate
turbulent and molecular viscosity. Using that scaling for the
viscosity, it is possible to build detailed and time dependent models
of accretion disks. For example, the disk surface density satisfies
the following diffusion equation as a result of mass and angular
momentum conservation
\begin{equation}
\frac{\partial \Sigma}{\partial t}=
\frac{3}{R}\frac{\partial}{\partial R}\left(R^{1/2}
\frac{\partial}{\partial R} \left( \nu R^{1/2} \Sigma\right)\right) \, .
\label{alpha_diffusion_eq}
\end{equation}
Such an equation can be used to calculate the time evolution of
$\Sigma$. It is not the purpose of this 
lecture, however, to provide an extensive review of $\alpha$--disk
models. The interested reader is referred to the classical textbook by
\citet{fkr_book} that exposes the details of such models. In the
introduction of this lecture, we shall content ourselves with an order
of magnitude estimate of $\alpha$. 

\begin{figure}
\begin{center}
\includegraphics[scale=1.2]{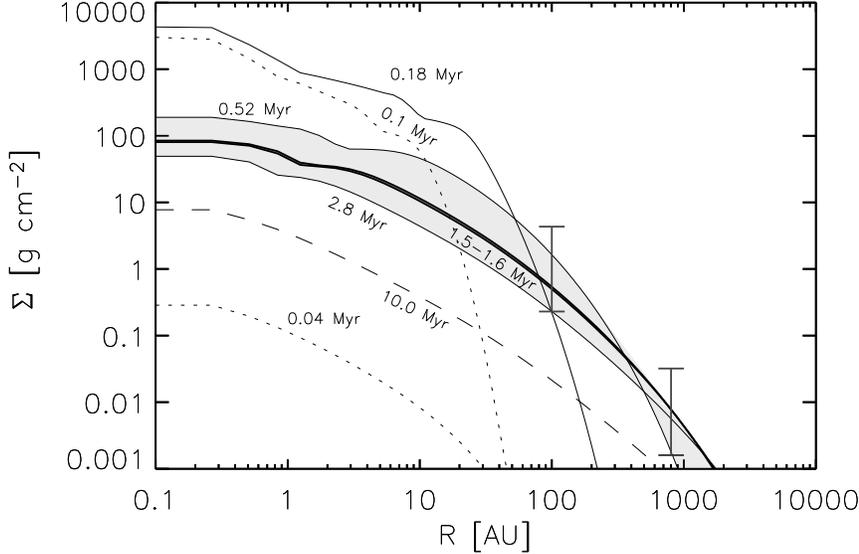}
\caption{Radial profile of the surface density obtained when solving
the diffusion equation given in a typical $\alpha$ disk model. The
different lines correspond to different times during the disk 
evolution, the vertical errors bars show the contraints provided by
observations of DM Tau and the grey shaded area shows the ensemble of
models that fit those constraints. Figure extracted from
\citet{hueso&guillot05}, to which the reader is referred for
further details on the model parameters.}
\label{hueso&guillot_fig}
\end{center}
\end{figure}

\noindent
What range of possible values can we expect? There are basically two
constraints. The first comes from the fact that turbulent eddies
should have a typical scale $l_t$ smaller than $H$ and that  
motions should be subsonic (i.e. the velocity fluctuations $\delta v_t
\le c_s$)\footnote{Indeed, any supersonic motion would quickly be
damped because of shocks. Likewise, turbulent structures at scales
larger than H would produce supersonic differential motions because
of the background shear}. As seen above, molecular viscosity can be
estimated as the product between the typical velocity of the molecules
and the typical distance they travel between collisions. By analogy,
turbulent viscosity can be estimated to be of order $l_t \delta
v_t$. The constraints on $l_t$ and $\delta v_t$ suggest an upper bound
of order $c_s H$ for the turbulent viscosity. In other words, we
expect $\alpha \le 1$. A second constraint on $\alpha$ comes from the
typical evolution timescale $t_{\textrm{PP}}$ of PP
disks. Observations suggest that $t_{\textrm{PP}} \sim 10^{6-7}$ years
\citep{armitage11}. Using the same reasoning as for molecular
viscosity, we can write
\begin{equation}
t_{\textrm{PP}} \sim \frac{R_d^2}{\nu_t} \, ,
\end{equation}
where $R_d \sim 100$ AU is now the disk outer radius. Using
$\nu_t=\alpha c_s H$ and $c_s=H\Omega$, we find\footnote{In a locally
  isothermal disk (i.e. the gas temperature is independent of the
  distance $Z$ to the disk midplane), hydrostatic equilibrium in the
  vertical direction writes $\partial P/\partial
  Z=-\rho GM/(R^2+Z^2)^{3/2}$. When $Z \ll R$ and using $P=\rho
  c_s^2$, one finds by integration that $\rho \propto \exp
  \left(-Z^2/2 H^2 \right)$, with $H=c_s/\Omega$ and
  $\Omega=\sqrt{GM/R^3}$ is the gas midplane angular velocity.}
\begin{equation}
\alpha \sim \frac{1}{2 \pi} \left( \frac{R}{H} \right)^2 \left(
\frac{T_{orb}(R=R_d)}{t_{PP}} \right) \sim 10^{-2} \textrm{ to }
10^{-1} \, .
\end{equation}
This argument can be carried a bit further as done for example by
\citet{hueso&guillot05}. These authors solved the diffusion equation
for the disk surface density given by Eq.(\ref{alpha_diffusion_eq}),
adding a source term on the right hand side of the equation. Such a term
is meant to model the infall of material from the envelope out of
which the protostar formed. The disk midplane temperature (required to
evaluate the turbulent viscosity) is calculated by including the
effect of turbulent heating, irradiation from the 
central star and radiative cooling from the disk surface. Using
reasonable initial conditions, they calculated the time evolution of
the radial profile of the surface density and compared their results
with observational constraints for two objects. The typical results
they obtained are illustrated on figure~\ref{hueso&guillot_fig}. To
account for 
millimeter observations of DM Tau, \citet{hueso&guillot05} concluded
that $\alpha$ should be in the range $10^{-3}$ to $10^{-1}$. Similar
constraints were drawn for the star GM Aur. These values are in broad
agreement with the simple orders of magnitude described above. The
paper by \citet{hueso&guillot05} is interesting because it illustrates
the possible use of $\alpha$--disk models in PP disks. It also
highlights that even detailed comparison with the observations only
results in very loose constraints on the transport of angular momentum
in PP disks that are no better than order of magnitude estimates. This
is not only due to the limits of the observations themselves (limited
angular resolution, uncertainties in the age of the objects) but also
to the simplicity of the $\alpha$--model, which assumes spatial as
well as temporal constancy of $\alpha$ and reduces turbulence as a
simple viscosity. 

The lesson not to forget from this discussion is
that $\alpha$ disk models are rough large scale models that can at
best offer an order of magnitude estimate of accretion disks
properties. Using them beyond there domain of validity is dangerous
business. They can serve as useful guides, especially when it comes to
comparing theoretical expectations with observations, but
understanding the underlying mechanisms that power the turbulence is
mandatory if we want to address all aspects of accretion disk
dynamics and angular momentum transport. This lecture is designed to
explore the properties of the most popular of these mechanisms, namely
the magnetorotational instability.

\paragraph{Outline of the lecture} The rest of the course is divided
in three parts. The first part focuses on the 
linear instability. A simple derivation is provided that is used to
extract the key properties of the MRI and to describe its physical
mechanism. The particular problems that arise in the case of
PP disks are emphasized. The second part is a tour of twenty years
of numerical simulations of the MRI. The techniques, results and
limits of those simulations are discussed. Finally, the last part
highlights a set key aspects of planet formation theories that
interface with the MRI. Outstanding issues are highlighted and
discussed along the way. 

\section{The Magnetorotational Instability}
\label{mri_lin_sec}

The goal of the present section is essentially to present a simple
derivation of the MRI. It will serve to establish the basic properties
of the instability. Despite the simplifications adopted in this section
(geometry of the field, locality of the analysis, over-simplified
treatment of the energetics), most of its conclusions are general.

\subsection{Governing equations}
\label{gov_eq_sec}

\begin{figure}
\begin{center}
\includegraphics[scale=0.3]{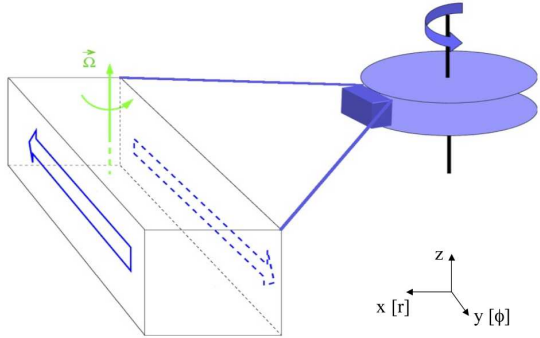}
\caption{Cartoon illustrating the local approach and coordinate system
  adopted for the derivation of the MRI dispersion relation
  (section~\ref{mri_lin_sec}) and for shearing boxes numerical
  simulations (section~\ref{mhd_turb_sec}). Figure courtesy
  G.~Lesur.}
\label{sb_pyl_fig}
\end{center}
\end{figure}

We begin by deriving a simplified form of the dynamical
equations. Our starting point is the equations of ideal MHD (see
chapter 1). Several approximations are made that
significantly simplify the derivation of the MRI dispersion relation.

\noindent
First, we assume the equation of state is isothermal: $P=\rho c_0^2$,
where $c_0$ is the constant sound speed of the gas. This removes the
energy equation from the problem. We next define a Cartesian
coordinate system centered at a radial distance $R=R_0$ from the
central mass and rotating around the central star with angular
velocity $\Omega_0$. The x--axis points in the radial direction, the
y--axis is directed along the direction of the flow rotation while the
z--direction is perpendicular to the equatorial plane of the disk (see
figure~\ref{sb_pyl_fig}). The associated unit vectors are noted
($\bb{e_x}$,$\bb{e_y}$,$\bb{e_z}$). We next consider a small volume
within the disk. For all points within that domain, we thus have in
particular $x \ll R_0$. In addition to pressure and the Coriolis
force, each fluid element is subject to the gravitational and
centrifugal accelerations. In the former, we neglect the vertical
component of the force because of the box small size (in practice, it
means we ignore the vertical density stratification of the disk). The
combined 
effect of gravitational and centrifugal acceleration can be expressed
as that of an equivalent tidal potential $\phi_{\textrm{eff}}$. If we
assume that the gas angular velocity follows a power law, $\Omega
\propto R^{-q}$, the tidal potential satisfies the
relation\footnote{For Keplerian flow, of course, $q=3/2$. We don't
specialize to that case in the following, though, as we want to
examine the stability of the flow as a function of its angular
momentum radial profile.} 
\begin{equation}
\del \phi_{\textrm{eff}}=-R\Omega^2 \bb{e_x} + R\Omega_0^2
\bb{e_x} = 2q\Omega_0^2x \bb{e_x}
\end{equation}
where we have used $R=R_0+x$ and a Taylor expansion that results from
the smallness of the domain. Straightforward integration yields:
\begin{equation}
\phi_{\textrm{eff}}=q\Omega_0^2x^2 \, .
\end{equation}
We are now in a position to write the equations governing the
fluid evolution. These equations are simply the MHD equations
discussed in the first chapter of this book, augmented with the
effective potential $\phi_{\textrm{eff}}$ and the Coriolis
force. Neglecting all dissipative terms (i.e. we work in the ideal MHD
limit), we obtain: 
\begin{eqnarray}
\frac{\partial \rho}{\partial t} + \del \bcdot (\rho \bb{v})  &=&  0
\,\label{continuity_eq} , \\
\rho \left[ \frac{\partial \bb{v}}{\partial t} +  (\bb{v} \bcdot \del )
\bb{v} \right] + 2 \rho \bb{\Omega_0} \btimes \bb{v}     &=&  - \bb{\nabla} P + \frac{1}{4\pi} (\del \btimes \bb{B})
\btimes \bb{B} + 2 q \rho
\Omega_0^2 x \bb{e_x} \label{momentum_eq} \\
\frac{\partial \bb{B}}{\partial t}  &=&  \del \btimes ( \bb{v} \btimes
\bb{B}  ) \, \label{induction_eq} .
\label{shearing_sheet_eq}
\end{eqnarray}
As usual, we have adopted the following notations: $\rho$ stands for
the gas density, $\bb{v}$ for the velocity, $\bb{B}$ for the magnetic
field, $P$ for the pressure, while $\bb{\Omega_0}=\Omega_0
\bb{e_z}$.

\subsection{Dispersion relation: a simple derivation}
\label{mri_disp_sec}

We now consider the simplest possible case where the system is
threaded by a uniform vertical magnetic field of strength $\bb{B}=B_0
\bb{e_z}$. In equilibrium, the gas density is uniform: $\rho=\rho_0$. 

\noindent
In such a situation, the equilibrium state is a balance
between the Coriolis force and centrifugal acceleration. It leads to
$v_y=-q \Omega x$. There is a linear shear. The stability of that
equilibrium is studied by considering small perturbations. For
simplicity, we consider here perturbations that only depend on $z$ and
$t$. All variables will thus be the sum of the equilibrium and time
independent part (denoted with a ``$0$'' subscript and the
perturbation (denoted with a prime exponent). For example, the density
writes: 
\begin{equation}
\rho = \rho_0 + \rho'(z,t) \, .
\end{equation}
The linearized equations on the perturbed density and vertical
velocity decouple from the other equations and write:
\begin{eqnarray}
\frac{\partial \rho'}{\partial t} + \frac{\partial}{\partial z}
(\rho_0 v_z') &=& 0 \, , \\
\rho_0 \frac{\partial v_z'}{\partial t} + c_0^2 \frac{\partial
  \rho'}{\partial z} &=& 0 \, , 
\end{eqnarray}
These two equations correspond to standard sound waves
propagating in the vertical direction. They are not affected by the
background rotation. The $z$ component of the induction equation
simply gives:  
\begin{equation}
\frac{\partial B_z'}{\partial t}=0 \, .
\end{equation}
Together with the condition $\bb{\nabla} \bcdot \bb{B}=0$, which
translates to the vertical derivative of $B_z'$ vanishing, this means
that $B_z'$ is a constant that can be incorporated into $B_0$.

\noindent
The remaining four equations ($x$ and $y$ components of the momentum
and induction equations) give:
\begin{eqnarray}
\rho_0 \frac{\partial v_x'}{\partial t} &=& -c_0^2 \frac{\partial
  \rho'}{\partial z} \\
\rho_0 \left[ \frac{\partial v_y'}{\partial t} - q \Omega v_x'  \right] &=& B_0 \frac{\partial B_y'}{\partial z} - 2
\rho_0 \Omega v_x' \\
\frac{\partial B_x'}{\partial t} &=& B_0 \frac{\partial v_x'}{\partial
  z} \\
\frac{\partial B_y'}{\partial t} &=& -q \Omega B_x' + B_0 \frac{\partial v_y'}{\partial
  z} \, .
\end{eqnarray}

\noindent
The set of equations derived above is homogeneous in $z$ and
$t$. This means we can take the Fourier transform by writing all
variables as:
\begin{equation}
X'=\overline{X'} \exp (i(\omega t - kz)) \, .
\end{equation}
Upon adopting this definition, the four equations describing
horizontal motions and magnetic field fluctuations then lead to the
following algebraic expressions:
\begin{eqnarray}
i \omega \rho_0 v_x' &=& -ik B_0 B_x' \\
i \omega \rho_0 v_y' &=& q \rho_0 \Omega v_x' -ik B_0 B_y' - 2 \rho_0
\Omega v_x' \\
i \omega B_x' &=& -ik B_0 v_x' \\
i \omega B_y' &=& -q \Omega B_x' - ik B_0 v_y'
\end{eqnarray}
where the overlines in the above expressions have been dropped for
clarity. Combining these equations leads after some algebra to the
following quartic equation that constitutes the dispersion relation of 
the problem:
\begin{equation}
\omega^4 - \omega^2 \left[ 2 k^2 v_A^2 + \kappa^2 \right] + k^2 v_A^2
\left[ k^2 v_A^2 - 2 q \Omega^2 \right] = 0 \, ,
\label{disp_mri_eq}
\end{equation}
where $\kappa^2=2(2-q)\Omega^2$ is the square of the epicyclic
frequency and $v_A=B_0/\sqrt{\rho_0}$ the Alfven velocity.

\noindent
In the absence of a magnetic field, $v_A=0$ and the dispersion
relation reduces to:
\begin{equation}
\omega^2 - \kappa^2 = 0 \, .
\end{equation}
Thus, the stability condition in hydrodynamic disks is $\kappa^2 \geq
0$, or equivalently $q \leq 2$. This is the Rayleigh criterion:
accretion disks with angular momentum increasing outward are linearly
stable. 

\noindent
The presence of a magnetic field dramatically modifies that classical
result. The sum $s$ of the two roots $\omega^2_\pm$ of
Eq.~(\ref{disp_mri_eq}) is given by $2s=2 k^2 v_A^2 + \kappa^2$. It is
always positive, which means that their product determines the stability
condition:
\begin{equation}
k^2 v_A^2 - 2 q \Omega^2 \geq 0 \, .
\end{equation}
The stability condition is thus $q \leq 0$. In the opposite case, $q
\geq 0$, there is always a critical value $k_c^2=2q\Omega^2/v_A^2$ of
the wavenumber such that $k^2 v_A^2 - 2 q \Omega^2$ is negative for
all $k<k_c$. Because of the definition of $q$, the system is unstable
whenever the angular velocity decreases outward, a condition that is
readily satisfied in accretion disks. A remarkable feature of the MRI,
first noted by \citet{chandrabookhydro}, is that the condition for
the flow to be destabilized by the magnetic field does not depend on the
strength of the field itself. This is because the Lorentz force
depends on the variations of the magnetic field along each field
line: the small scale oscillations of a small B--field results in the
same Lorentz force as the large scale oscillation of a large field
\citep{balbus&hawley91}. 

\subsection{MRI properties}

\paragraph{The fastest growing mode} Using
equation~(\ref{disp_mri_eq}), the most unstable mode is characterized
by a growth rate $\sigma_{\textrm{max}}$ and a wavenumber $k_{\textrm{max}}$ that
satisfy the following relations: 
\begin{eqnarray}
\sigma_{\textrm{max}} &=& \frac{q \Omega}{2} \\
k_{\textrm{max}}^2 v_A^2 &=& \frac{q}{4}(4-q)\Omega^2 \, .
\end{eqnarray}
For accretion disks in Keplerian rotation, $q=3/2$ which results in
$\sigma_{\textrm{max}}=0.75 \Omega$ and $k_{\textrm{max}} v_A \sim \Omega$. As noticed
early on \citep{balbus&hawley91,balbus&hawley98}, this is 
an enormous growth rate as it results in amplification factors in
energy per dynamical time larger than $10^4$
\citep{balbus&hawley98}. In fact, \citet{balbus&hawley92a} conjectured
that the MRI growth rate is the fastest growth rate for instabilities
resulting from differential rotation. This is one of the reason the
MRI quickly became important in dynamical studies of accretion
disks. 

\noindent
The structure of the most unstable mode of the MRI consists in horizontal
layers of alternating radial velocities and radial magnetic
fields that grow in amplitude. The action of shear means that
azimuthal velocities and magnetic fields are produced. Vertical
velocity and density 
perturbations both vanish (which means that the most unstable mode of
the MRI is incompressible). These growing ``fingers'' of perturbed
magnetic fields and velocities are now known as channel modes. 

\paragraph{A weak field instability} The instability criterion derived
above states that all modes of wavelength larger than a critical scale
(given by the wavenumber $k_c$) are unstable. Of course, these
unstable modes cannot be larger in scale than the typical vertical
size of the system. A necessary condition for the MRI to operate is
thus that the smallest of those modes fit within the disk:
\begin{equation}
k_c \geq \frac{2\pi}{H}
\end{equation}
This equation can be written as a condition on the plasma parameter
$\beta$, the ratio between thermal and magnetic pressure:
\begin{equation}
\beta \geq \frac{4\pi^2}{q} \, .
\end{equation}
Since the parameters entering the right hand side of this relation are
all of order unity, this is a requirement that the magnetic field
strength should be smaller than a given threshold for which magnetic
and thermal energy are roughly in equipartition. This is why the MRI
is often qualified as being a weak field instability.

\begin{figure}
\begin{center}
\includegraphics[scale=0.48]{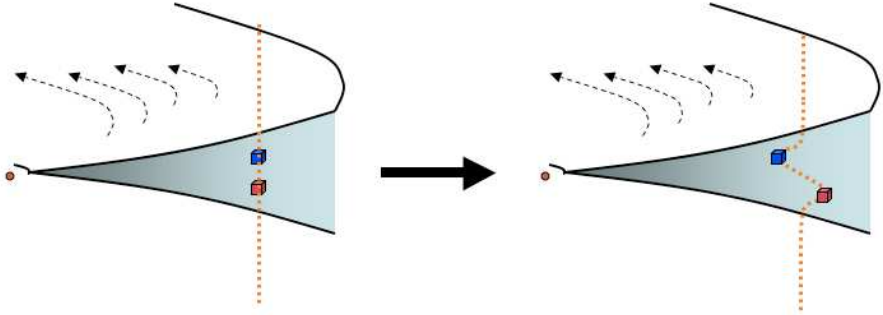}
\caption{Cartoon illustrating physical mechanism of the MRI (the black
dashed line shows the radially decreasing angular velocity of the
gas). Consider two fluid elements at the same cylindrical radius
threading a weak vertical field line (shown with the red dotted line),
then displaced as shown in the figure. To maintain their 
respective angular momenta, the (upper) inward moving mass rotates
more rapidly and the (lower) outward moving mass less rapidly. This
creates an azimuthal tension force in the field line that transfers
angular momentum from the inwardly displaced mass to the outwardly
displaced mass.  Since the inward mass loses angular momentum it
drifts further inward; the outward mass, acquiring angular momentum,
moves further outward. This increases the field line tension and the
process runs away.}
\label{mri_mechanism_fig}
\end{center}
\end{figure}

\paragraph{Physical interpretation} The physical mechanism of the
instability is illustrated in figure~\ref{mri_mechanism_fig}. Consider
two fluid elements sitting on a particular magnetic field line that is
purely vertical when unperturbed. When that field line is perturbed,
these two fluid elements are displaced and follow the field line
perturbation (this is because, in the ideal MHD limit, fluid particles
and magnetic fields are tied together). They move to different radial
location. In order to conserve their angular momenta, they start to
rotate at different angular velocity and begin drifting apart. This drift 
stretches the magnetic field line that link the two fluid particles and
results in a magnetic force (the increased tension of that field lines)
on both fluid elements. That force slows (accelerates) the inner
(outer) fluid element which moves further inward (outward) in a
runaway process. In the meantime, angular momentum has been exchanged
between them.

\paragraph{A robust instability} The situation considered above was
deliberately chosen to be oversimplified in order to highlight the
basic properties of the instability and the physical mechanism at
play. Focus has been placed on the simplest perturbations to 
highlight the physical property of the instability. Axisymmetry has been
assumed and we have considered perturbations that do not depend on $x$. In
fact, there exists other modes with a finite $k_x$
\citep{balbus&hawley91} as well as non-axisymmetric modes
\citep{hawley&balbus92}. They have smaller growth rates than the
channel modes discussed above. In addition, the MRI grows in
conditions that are more general than the simple setup considered
above. the MRI grows in the presence of a
radial magnetic field \citep{balbus&hawley91} or a pure azimuthal
magnetic field \citep{hawley&balbus92}. In this latter case, the
perturbations are required to be non axisymmetric, which renders the
analysis quite complicated because of the background shear, and the
growth is only transient for finite vertical wavenumbers. Normal modes
also exist when the vertical density stratification of the disk is
accounted for \citep{gammie&balbus94}. To state things simply, the MRI
grows as long as a weak magnetic field is present in a sheared
rotating flow where angular velocity decreases outward.

\subsection{The case of protoplanetary disks}

As described above, the MRI is a powerful linear instability that
displays enormous growth rates in the ideal MHD limit. However, an
implicit assumption of the above analysis is that there is good 
coupling between the gas and the magnetic field. We have assumed that
the gas is a perfect conductor. However, in cold and dense
environments such as in PP disks, this is not the case. We will see
below that simple estimates of the electron fraction in PP disks confirm
that this can significantly alter the MRI properties describing
above. Before doing so, we focus on the requirement for the MRI to 
operate when the effect of dissipation coefficients is taken into
account. 

\paragraph{The effect of diffusion coefficients}

The MRI dispersion relation was derived above in the limit of
dissipationless MHD. As mentioned already, such an assumption is not
realistic because of the large plasma diffusivity. This is why various
authors have explored the influence of a finite resistivity and viscosity 
\citep{lesur&longaretti07,pessah&chan08}. The results of such studies
is not surprising: dissipation tends to reduce the growth rate of the
instability. The detailed modifications on both the growth rate and
the eigenmodes can be analyzed in details but turns out to be
complicated. However, it is possible to use simple 
scaling arguments that illustrate their importance for PP disks. Here
we follow the discussion of \citet{flemingetal00} who focused on the
case where dissipation is dominated by a large ohmic resistivity
$\eta$. In such a situation, the induction equation given by
Eq.~(\ref{induction_eq}) is modified and writes
\begin{equation}
\frac{\partial \bb{B}}{\partial t}  =  \del \btimes \left( \bb{v} \btimes
\bb{B} - \eta  \del \btimes \bb{B} \right) \, .
\label{induction_plus_ohmic_eq}
\end{equation}
The additional term in the induction equation is responsible for
diffusing the magnetic field. As a result, MRI modes with a wavenumber
$k$ are expected to be affected when their growth rate is of order
the diffusion rate $\eta k^2$ associated with that diffusivity. For
the most unstable MRI mode, the growth rate is of order $\Omega$ and
is thus affected by Ohmic resistivity when:
\begin{equation}
\eta k^2 \sim \Omega \, .
\end{equation}
In addition, its wavenumber $k$ roughly satisfies $k v_A \sim
\Omega$. Combining the two expression, we can expect the MRI to be
stabilized when 
\begin{equation}
\Lambda = \frac{v_A^2}{\eta \Omega} \le 1 \, .
\label{elsasser_eq}
\end{equation}
The dimensionless parameter $\Lambda$ is called the Elsasser
number. Of course, small scale modes (large $k$) are affected
first by the resistivity while larger and larger resistivity is
required to stabilize the large scale modes (small $k$) and the simple
argument above is only meant to give an order of magnitude
estimate. Nevertheless, the simple criterion that the Elsasser number
needs to be larger than unity for the MRI to operate is confirmed by
more detailed analysis. 
 
\noindent
Now since the resistivity is a decreasing function of electron
abundance, the above criterion also shows that a small electron fraction
will tend to stabilize the flow. Let us try to be more
quantitative. What is the typical electron fraction that is required
at 1 AU in a typical PP disk? To answer 
that question, \citet{blaes&balbus94} used standard expressions for
the conductivity, such as given by \citet{spitzer62} for typical
astrophysical plasmas. They provide a closed form for the resistivity in
a disk consisting of ions  and neutrals:
\begin{equation}
\eta= \frac{230 \, T^{1/2}}{x_e} \, \textrm{cm}^2 .\textrm{s}^{-1} \,
.
\label{ohmic_res_eq}
\end{equation}
In the following we consider typical values for the disk parameters at
1 AU, i.e. $c_s=1$ km.s$^{-1}$, $\Omega=2 \times 10^{-7}$ s$^{-1}$ and
a temperature of a few hundreds Kelvin. We can expect the magnetic
field to reach at most equipartition (an expectation confirmed by
numerical simulations), so that we take $v_A \sim c_s$. Combining
Eq.~(\ref{ohmic_res_eq}) with the requirement that $\Lambda$ should be
larger than unity, we see that the boundary between stable and
unstable flows lies at $x_e \sim 10^{-13}$. This is an extremely small
number, but we shall see in the following that there are indeed
regions of the disks where the electron fraction is much smaller.

\paragraph{PP disks are cold} 

\noindent
The simplest way to produce electrons in
gases is through thermal ionization produced by collisions between
particles. Temperature in PP disks range from a few thousands Kelvins
in their inner regions to a few tens of Kelvins. The number of electrons
produced by thermal ionization in such circumstances can be worked out
using the Saha equation. For atoms with a single level of ionization,
it writes \citep{spitzer62} 
\begin{equation}
\frac{n_e^2}{n_i}=\frac{2}{\Lambda_B^3}\frac{g_1}{g_0} \exp \left( -
\frac{\epsilon}{k_B T} \right) \,\,\,  \textrm{where} \,\,\,
\Lambda_B=\sqrt{\frac{h^2}{2 \pi m_e k_B T}}
\label{saha_textbook}
\end{equation}
is the thermal de Broglie wavelength, which amounts to $\sim 2.4
\times 10^{-7} \, \textrm{cm}$ when $T=10^3$ K. The other terms are
the density of electrons $n_e$ and ions 
$n_i$, their respective statistical weights $g_1$ and $g_0$, the
ionization energy $\epsilon$, the Boltzmann constant $k_B$, the Planck
constant $h$ and the electron mass $m_e$. The main producers of
electrons are species having the lowest ionization potential. In PP
disks, such species are essentially alkali atoms (Sodium and
Potassium), for which $\epsilon$ is of order $5 \,
\textrm{eV}$. Equation~(\ref{saha_textbook}) can be 
written in term of their abundance relative to hydrogen. This gives
the following expression for the electron fraction $x_e=n_e/n_H$:
\begin{equation}
x_e=\frac{\sqrt{2}}{\Lambda^3n_H}a^{1/2}\frac{g_1}{g_0} \exp \left( -
\frac{\epsilon}{2 k_B T} \right) \, ,
\end{equation}
where $a=n_K/n_H \sim 10^{-7}$ is the abundance of alkali atoms
\citep{umebayashi&nakano88}. The important point to notice in the
above equation is the overwhelmingly importance of the exponential
\citep{balbus&hawley00}. Indeed, the term $\exp
\left(-\epsilon/2 k_B T \right)$ respectively amounts to about $5
\times 10^{-7}$, $3 \times 10^{-13}$ and $7 \times 10^{-26}$ when
$T=2000$, $1000$ and $500$ K. This is a variation by almost $20$
orders of magnitudes while the temperature only varies by a factor of
four! Over the same range of temperature, the term
multiplying the exponential only varies by a factor of a few. As a
consequence of that exponential factor, the electron fraction $x_e$
equals $5 \times 10^{-7}$, $10^{-13}$ and $10^{-26}$ respectively when
$T=2000$, $1000$ and $500$ K. This simple estimate along with the
considerations of the above section suggest that the MRI will be
strongly affected by non-ideal MHD effects once the temperature drops
below $1000$ K. In PP disks, this corresponds to all locations beyond
a few tens of a AU from the central star \citep{dalessioetal98}.

\paragraph{PP disks are dense}

\noindent
The second reason for the small electron fraction in PP disks comes
from their large gas density. The midplane number density of hydrogen
molecules is of the order of $10^{14}$ cm$^{-3}$ (see
section~\ref{intro_sec}), which should be compared to $\sim 1$
cm$^{-3}$ for the diffuse ISM or $\sim 10^5$  cm$^{-3}$ for dense star
forming cores \citep{maclow&klessen04}. This means that PP disks can
hardly be ionized by external sources, like UV and X-ray photons or
cosmic rays. Again, this can be recovered by the following simple
calculation. Consider for simplicity the case of a pure gas disk
(i.e. no dust particles) irradiated by a flux of cosmic rays. Free
electrons are produced from molecular hydrogen at a 
rate $\xi$:
\begin{equation}
H_2 \rightarrow H_2^+ + e^-
\end{equation}
As discussed by \citet{oppenheimer&dalgarno74}, the positively charged
molecules $H_2^+$ reacts almost instantaneously with the numerous
neutral $H_2$ molecules to produce a series of molecular ions like
$H_3O^+$, $HCO^+$, $O_2^+$, $H_3^+$. These ions
(collectively denoted $m^+$ in the following) dissociatively
recombines with the free electrons with a reaction rate $\beta$:
\begin{equation}
m^+ + e^- \rightarrow m \, .
\end{equation}
These simple reactions translates into the following differential
equation that describes the time evolution of the molecular ion number
density: 
\begin{equation}
\frac{dn_{m^+}}{dt}=\xi n_{H_2} - \beta n_{m^+} n_{e^-}
\label{balance_ionize_eq}
\end{equation}
Assuming electroneutrality in the fluid (i.e. $n_{m^+}=n_{e^-}$) and
steady state gives
\begin{equation}
x_e=\sqrt{\frac{\xi}{\beta n_{H_2}}} \, .
\label{ionization_rate_CR}
\end{equation}
The reaction rate for dissociative recombination is also given by
\citet{oppenheimer&dalgarno74}: $\beta \sim 3 \times 10^{-6}
T^{-1/2} \sim 10^{-7}$ cm$^3$.s$^{-1}$ for a typical temperature of a
few hundreds Kelvin. The ionization rate value is more debated. It is
well known that galactic cosmic-rays have difficulties entering the
inner solar system because of the solar wind \citep{gammie96}. The
same is probably true of other young stars. At the same
time, recent work suggests that the ambient ionization rate by cosmic
rays in clusters (i.e. at the birth place of most stars) could be
larger by three orders of magnitudes \citep{fatuzzoetal06}. A
conservative estimate of the cosmic rays ionization rate based on
present day measured values gives 
$\xi=10^{-17}$ s$^{-1}$. When used in Eq.~(\ref{ionization_rate_CR}),
one finds $x_e \le 10^{-12}$ (the upper limit comes from the fact that
the cosmic ray ionization rate is much reduced at the disk
midplane, so our estimates is in fact an upper limit of the electron
fraction). Again, this shows that the electron fraction is very small
in PP disks and we can therefore anticipate that only the disk surface
layers are sufficiently ionized for the MRI to become active.
 
\paragraph{The dead zone paradigm}

\begin{figure}
\begin{center}
\includegraphics[scale=0.3]{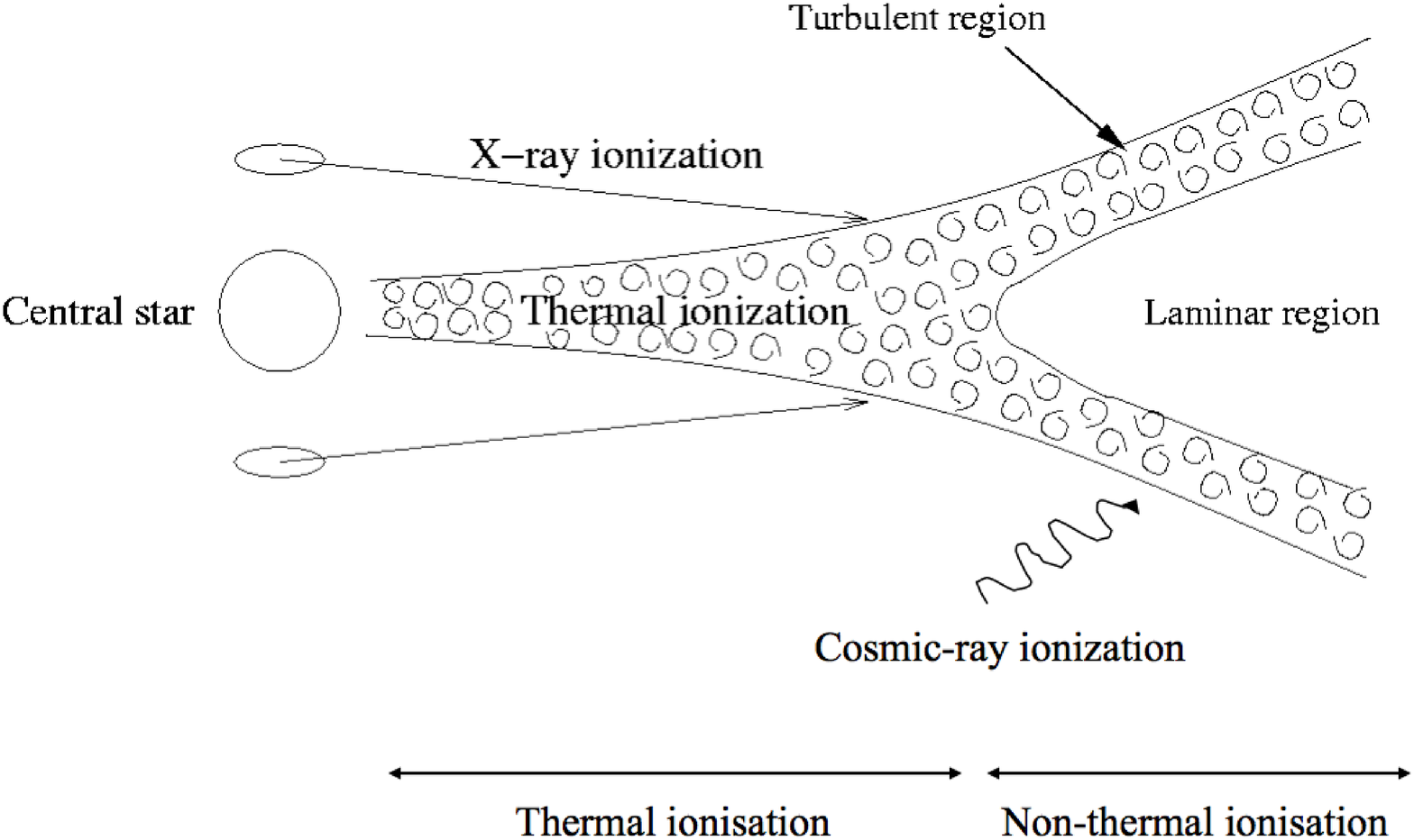}
\caption{Cartoon illustrating the state of PP disks as suggested by
  \citet{gammie96}. The innermost parts are MRI unstable as a result of
  thermal ionization which brings temperature above $10^3$ K. In the
  disk inner parts, $0.1$--$1<R<10$--$20$ AU, the disk is too cold for
  thermal ionization to be efficient and too massive for nonthermal
  ionization sources to produce enough free electrons all the way down
  to the disk equatorial plane. The MRI develops only in the disk
  upper layers where the ionization fraction is large enough as a
  result of X-ray and/or cosmic rays ionization.}
\label{flow_prop_fig}
\end{center}
\end{figure}

The simple calculations described above resulted in the dead zone
paradigm proposed 
by \citet{gammie96}: the MRI can only develop in regions that are either hot
enough ($T>10^3$ K) for thermal ionization to occur or in regions
where cosmic rays (or other nonthermal ionization sources) can
penetrate. In PP disks, the former condition is satisfied only close to the
central star, at stellocentric distances smaller than a few tens of an
AU. The second region (where nonthermal ionization is sufficient to
activate the MRI) forms a thin layer at the disk surfaces. The MRI is
expected to grow in both regions (and, as we shall see in the
remaining of this lecture, the flow becomes turbulent there). In the
disk midplane, the MRI is quenched and the flow remains laminar. The
structure of the disk that results is illustrated in
figure~\ref{flow_prop_fig}. It can immediately be seen that 
such a disk structure is qualitatively different from standard
$\alpha$ disk models. Although this has recently been challenged
\citep{terquem08}, it was for example noted early on
that such PP disks would have trouble evolving toward a steady
state \citep{gammie96}. This is because the accretion rate in the disk
inner parts would be larger than the accretion rate that those active
layers at the disk surfaces could sustained. This mismatch has led to a class
of disk models that display an eruptive behavior, sometimes aimed at
explaining the FU Orionis phenomenon
\citep{zhuetal10a,zhuetal10b,martinetal12}. 

\noindent
It should be emphasized that the issue described above is serious
and can compromise accretion onto young star in PP disks. The
active layers at the disk surface can be so thin that 
their ability to sustain the observed inward mass flux is
questionable. In addition, the magnetic field strength can reach
equipartition at those dilute locations, stabilizing the MRI even if
ideal MHD conditions prevail. These are the reasons why people have
built detailed chemical models of disks aimed at a precise
determination of the electron fraction. Additional physical processes
have been considered as potential solution to the problem, like X-ray 
irradiation \citep{glassgoldetal97}, 
radioactive decay \citep{turner&drake09}, far UV ionization
\citep{perezbecker&chiang11} or the effects of metal atoms 
\citep{fromangetal02}. The basic result that large portion of PP disks 
are immune to the MRI is robust. It becomes even worse when the
effects of dust grains are considered. In the presence of dust 
grain (labeled "gr"), electrons can recombine through the reaction
\begin{equation}
gr + e^- \rightarrow gr^- \, .
\end{equation}
The number of such reactions can be estimated as the product
between the grains cross section $\sigma=a^2$, the electrons thermal velocity
$u_e$ and the number density of grains (noted $n_{gr}$) and of electrons. When
taking that reaction into account, Eq.(\ref{balance_ionize_eq})
is modified and becomes, in steady state: 
\begin{equation}
\xi n_{H_2} - \beta n_{m^+} n_{e^-} - \sigma u_e n_{gr} n_{e^-} = 0 \,
.
\label{ionize_with_grains_eq}
\end{equation}
In the absence of grains, $n_{gr}=0$ and we recover the results of the
preceding section. The key aspect of that equation is that the third
term quickly dominates the second as soon as 
dust grains are introduced. If we assume that all the solids are in
grains of size $a=10 ~ \mu$m (in which case $\sigma \sim 10^{-6}$
cm$^2$) and that the dust-to-gas ratio is $10^{-2}$ in mass, then the
dust fraction $x_{gr}=n_{gr}/n_n \sim 10^{-12}$. For temperatures of
order $100$ K, we have $u_e \sim 10$ km.s$^{-1}$. Putting things
together, the ratio between the second and third terms of
Eq.~\ref{ionize_with_grains_eq} can be calculated: 
\begin{equation}
\frac{\beta n_{m^+}}{\sigma u_e n_{gr}} \sim 10^{-7}
\frac{x_{m^+}}{x_{gr}} \ll 1
\end{equation}
unless $x_{m^+}$ is larger than $10^{-5}$, i.e. if the medium is
completely ionized as far as the MRI is concerned. In the regime of
small ionization fraction, the electron fraction is thus given by a
balance between ionization and recombination on grains and writes: 
\begin{equation}
x_e=\frac{\xi}{\beta_{gr} n_{H_2}} \, .
\label{ionization_rate_elec}
\end{equation}
where we have introduced $\beta_{gr}=\sigma u_e x_{gr}=10^{-12}$
cm$^3$.s$^{-1}$. Using the same values as before for $\xi$ and
$n_{H_2}$ in that equation, one
obtains $x_e \sim 10^{-18}$. This is a significant reduction compared
to gas phase chemistry. Formally, it can be traced to the absence
of a square root in Eq.~(\ref{ionization_rate_elec}). This is because
ion/electron recombination requires electrons colliding with ions,
the number of which is proportional to that of electrons, while
electron/grain recombination requires electrons colliding with
grains, the number of which is independent to that of electrons. 

\noindent
The simple calculation above neglects several aspects of the problem,
among which charge balance and charged grains reaction. It also
assumes a single type of ions and a single grain size. It has
nevertheless been confirmed in the past few years by detailed numerical
integration of complex chemical networks 
\citep{ilgner&nelson06a,ilgner&nelson06b,bai11b}. The effect of dust grains
is a severe problem for angular momentum transport in PP disks.

\begin{figure}
\begin{center}
\includegraphics[scale=0.6]{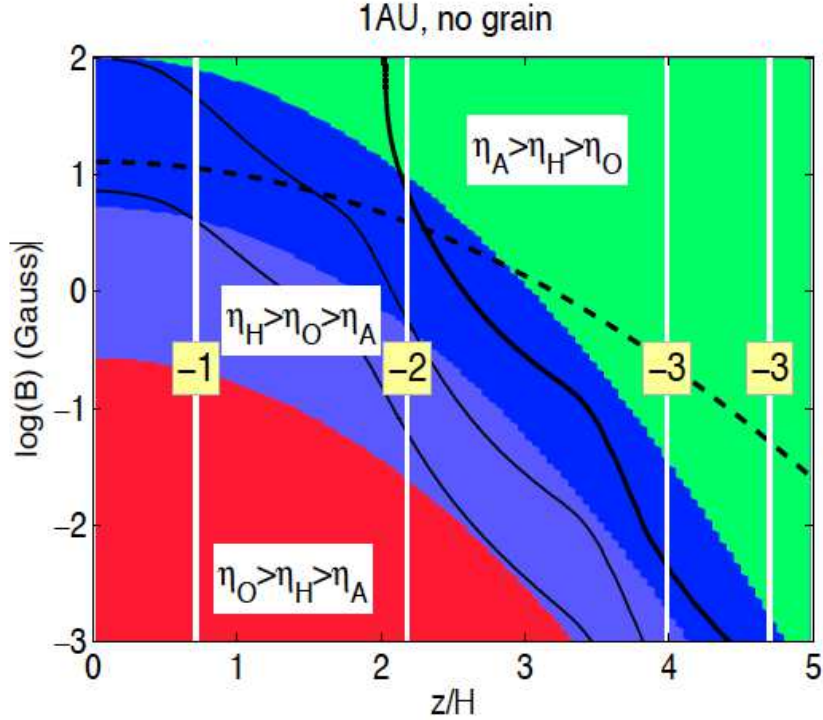}
\caption{Non--ideal MHD regimes (as a function of vertical position
and magnetic field strength) for a typical PP disk model at $1$ AU,
assuming all dust grains have grown to form large particules. The
red parts of the plot are dominated by Ohmic diffusion, blue parts are
dominated by the Hall term while green parts are regions of the disk
where ambipolar diffusion is the largest. From \citet{bai11a}.}
\label{bai11_fig}
\end{center}
\end{figure}

\paragraph{Ambipolar diffusion \& the Hall effect} In fact, things are
even more complicated than discussed above! This is because ohmic
diffusion is not the only important non--ideal MHD effect in PP
disks. Ambipolar diffusion and the Hall effect can be important as
well. In chapter 1 of the book \citep[see also][]{balbus11}, we
derived the induction equation including all non--ideal terms. In the
absence of dust grains, it writes 
\begin{equation}
\frac{\partial \bb{B}}{\partial t}  =  \del \btimes \left( \bb{v} \btimes
\bb{B} - \eta  \del \btimes \bb{B} - \frac{\bb{J} \btimes \bb{B}}{n_e
  e} + \frac{(\bb{J} \btimes \bb{B}) \btimes \bb{B}}{\gamma \rho_i
  \rho c} \right) \, ,
\label{induction_full_eq}
\end{equation}
where, in addition to ohmic diffusion already discussed, see
Eq.(\ref{induction_plus_ohmic_eq}), the second term accounts for the
Hall effect and the last term is due to ambipolar diffusion. It is
possible to rewrite that equation in a form that is easier to
interpret \citep{bai11a}:
\begin{equation}
\frac{\partial \bb{B}}{\partial t}  =  \del \btimes \left( \bb{v} \btimes
\bb{B} - \frac{4\pi\eta}{c}  \bb{J} - \frac{4\pi\eta_H}{c} \bb{J}
\btimes \bb{\hat{B}} - \frac{4\pi\eta_A}{c} \bb{J_{\perp}}  \right) \, , 
\label{induction_full_eq}
\end{equation}
where $\bb{\hat{B}}$ is a unit vector in the direction of $\bb{B}$ and
$\bb{J_{\perp}}$ is the component of $\bb{J}$ that is perpendicular to
$\bb{B}$. Order of magnitude estimates for $\eta_H$ and $\eta_A$ can
be obtained by comparing these two expressions:
\begin{eqnarray}
\eta_H &\sim& \frac{cB}{4\pi en_e} \label{resist_hall} \\
\eta_A &\sim& \frac{B^2}{4\pi \gamma_i \rho_i \rho} \label{resist_ambi} \, .
\end{eqnarray}
A similar expression for the Ohmic resistivity is provided by
Eq.(\ref{ohmic_res_eq}). The effect of ambipolar diffusion and the
Hall term on the MRI are not straightforward. The former has been
considered by \citet{blaes&balbus94} and the latter has been analyzed
later on by \citet{wardle99} and \citet{balbus&terquem01}. The effect
of the Hall term can be 
stabilizing or destabilizing depending on the relative orientation
between $\bb{\Omega}$ and $\bb{B}$. In the appropriate geometry, it is
so strong that it has the potential to overcome the stabilizing effect
of Ohmic diffusion \citep{wardle&salmeron12}. Despite that complexity,
one can guess from the form of the induction equation that the MRI is
stabilized when the following Elsasser numbers associated with the
effect of the Hall term or with that of ambipolar diffusion fall below
unity:
\begin{eqnarray}
\chi &=& \frac{v_A^2}{\eta_H \Omega} \\
Am &=& \frac{v_A^2}{\eta_A \Omega} \, .
\end{eqnarray}
The expression given by Eq.~(\ref{resist_hall}) and
(\ref{resist_ambi}) for the effect of the Hall and ambipolar
resistivities can help get a feeling for the relative importance of
both terms. In making these estimates, the strength of the magnetic
field is very uncertain. \citet{wardle97} suggests field strengths
that range between $10^{-3}$ Gauss and $10^2$ Gauss, and we shall take
the lower 
value of that range for illustrative puposes. Using values for the
neutral density typical of PP disks midplane at $1$ AU, $n_H \sim
10^{14}$ cm$^{-3}$, as well as $\gamma_i=2.8 10^{13}$
cm$^3$.s$^{-1}$.g$^{-1}$, $\rho_i=39 \rho$ \citep{balbus&terquem01}
and a typical temperature $T \sim 100$ K, we can write 
\begin{eqnarray}
\eta &\sim& \left( \frac{10^3}{x_e} \right) \textrm{ cm}^2\textrm{.s}^{-1} \, , \\ 
\eta_H &\sim& \left( \frac{50}{x_e} \right) \left( \frac{10^{14} \textrm{cm}^{-3}}{n_H}\right)
\textrm{ cm}^2\textrm{.s}^{-1} \, , \\
\eta_A &\sim& \left( \frac{2.6 \times 10^{-3}}{x_e} \right) \left( \frac{10^{14}
  \textrm{cm}^{-3}}{n_H} \right)^2 \textrm{ cm}^2\textrm{.s}^{-1} \, .
\end{eqnarray}
As can be seen, all of these resistivities scale like $x_e^{-1}$. In
PP disks midplane at $1$ AU, $n_{H_2} \sim 10^{14}$ cm$^{-3}$ and
ohmic resistivity dominates. The scaling 
$\eta_H \propto n_H^{-1}$ and $\eta_A \propto n_H^{-2}$ suggests that
the Hall term and ambipolar diffusion become more and more important
in the disk upper layers. Indeed, at three scaleheights above the
midplane \footnote{Assuming the gas vertical profile to be Gaussian,
  $n_{H_2}(Z=3H) \sim 1.1 \times 10^{-2} 
  n_{H_2}(Z=0)$}, we have $\eta_H \sim 4.5 \times 10^{3}/x_e$
cm$^2$.s$^{-1}$ and $\eta_A \sim 21/x_e$ cm$^2$.s$^{-1}$, thus $\eta_H
\ge \eta \ge \eta_A$. At five disk scaleheights\footnote{At $Z=5H$,
  $n_{H_2}(Z=5H) \sim 3.4 \times 10^{-6} n_{H_2}(Z=0)$}, we have
$\eta_H \sim 1.3 \times 10^{7}/x_e$ cm$^2$.s$^{-1}$ and $\eta_A \sim
1.9 \times 10^8/x_e$ cm$^2$.s$^{-1}$, thus $\eta_A \ge \eta_H \ge
\eta$. This simple discussion demonstrates that we can expect PP disks
midplane to be dominated by Ohmic resistivity, while the Hall effect
and ambipolar diffusion should successively dominates the disk
atmospheres. This order of magnitude estimate is confirmed by
detailed calculations of the electron fraction that take into account
a complex chemical network including dust grains. An example is given by
the results of \citet{bai11a} some of which are illustrated by
figure~\ref{bai11_fig}. It confirms the results anticipated above:
ohmic diffusion tends to dominate in the disk midplane, before being
successively replaced by the Hall effect and ambipolar diffusion in
the disk upper layers. Although figure~\ref{bai11_fig} is a plot of
the disk structure at $1$ AU from the central star, the outer disk
retain the same qualitative properties. We will come back to the
consequences of these non--ideal terms during the nonlinear evolution of
the MRI in section~\ref{saturation_mri_sec} and \ref{ppdisk_struct_sec}.

\section{MHD turbulence}
\label{mhd_turb_sec}

\begin{figure}
\begin{center}
\includegraphics[scale=0.5]{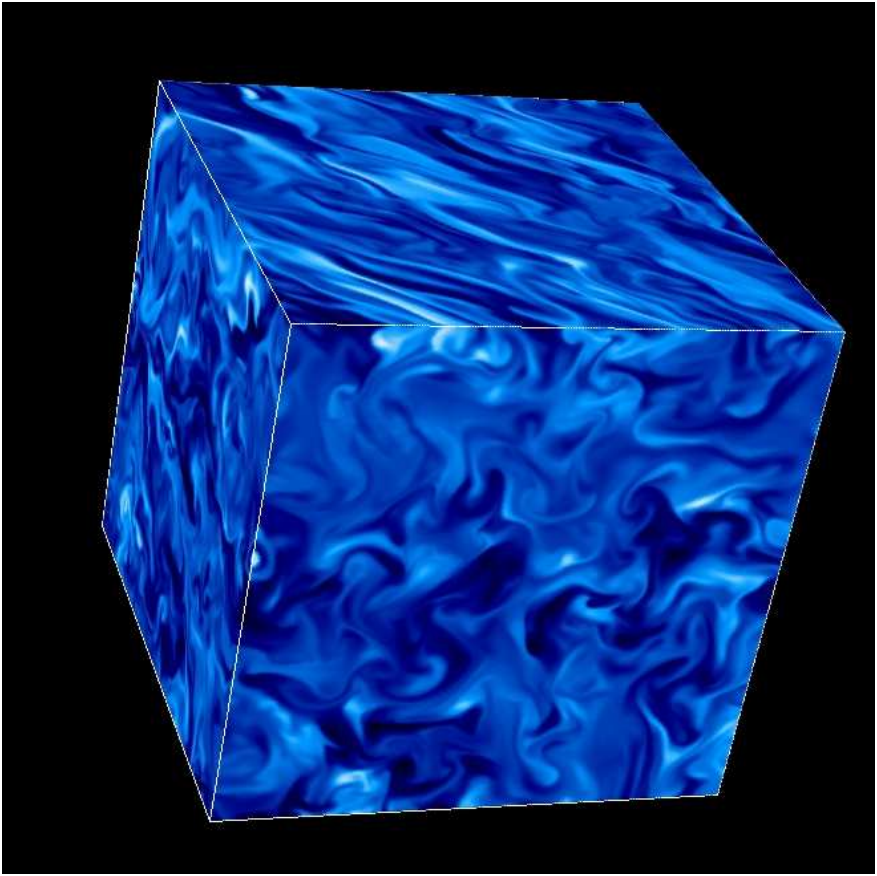}
\caption{Volume rendering illustrating the typical structure of the
  azimuthal component of the magnetic field in a typical MHD numerical
  simuations of the MRI performed in the idealized framework of the
  shearing box model (see section~\ref{mhd_turb_sec} for details). The
  flow is turbulent.}
\label{highres_sim_fig}
\end{center}
\end{figure}

So far, we have focused only on the linear aspects of the MRI. But of
course, as it grows, nonlinear terms in the MHD equations 
start to influence the flow. Its properties during that stage are best
studied with the help of numerical simulations. In this section, we
review the results that have been established using such simulations
in the last twenty years. The focus of this section is on idealized
numerical experiments such as shown in
figure~\ref{highres_sim_fig} that aim at understanding MHD turbulence
as an angular momentum transport process. This is a neccessary first
step before building realistic models of protoplanetary disk structure
and examining the consequences of the MRI for planet formation models.

\subsection{Methods and early results}
\label{turb_history_sec}

The simultaneous discovery of the MRI and the rise in computing power 
witnessed during the early 90's quickly established a series of
results that demonstrated the tremendous potential of that
instability. In this section, we briefly describe the methods that
were developed to that end before summarizing the most important of
these findings.

\paragraph{The shearing box model} 

\begin{figure}
\begin{center}
\includegraphics[scale=0.35]{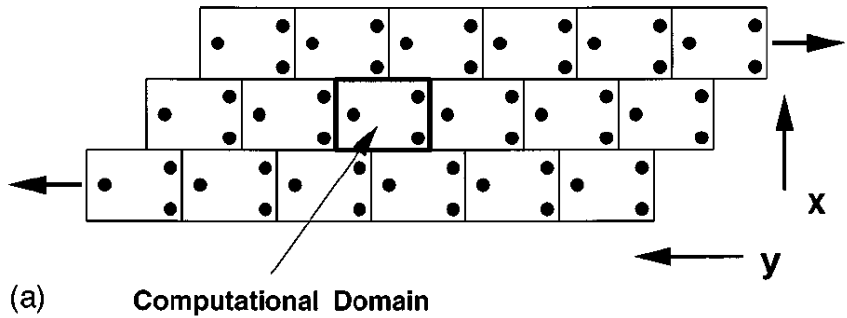}
\caption{Cartoon illustrating the origin of the shearing box radial
boundary conditions. The idea is that the disk is composed of many
such boxes, statistically identical, that are sliding past each other
because of the background orbital shear. Adapted from
\citet{balbus&hawley98}.} 
\label{shbc_fig}
\end{center}
\end{figure}

As discussed in section~\ref{gov_eq_sec}, the physics of the MRI is
easier to understand if the analysis is made local. That local
approach has proved very useful ever since the discovery of the
MRI. It is now known as the shearing box model. It helps to focus on
the dynamics of the flow without the burden of having to consider the
entire disk (with additional difficulties such as ill--posed boundary
conditions for example). This is of course at the cost  
of the realism of the approach, but the shearing box model has led to
spectacular progress (and still does!). What is true for the
analytical analysis remains true when developing numerical simulations
of the MRI. The set of equations that is used in this case is derived
from the MHD equations as described in section~\ref{gov_eq_sec}. There
are two possible variants: one is to stick to the analysis detailed
there and solve Eq.~(\ref{continuity_eq}), (\ref{momentum_eq})
and (\ref{induction_eq}). But, as we have seen, the fluid perturbations
associated with the channel modes of the MRI are incompressible. An
alternative is thus to use the incompressible MHD equation (the
continuity equation then reduces to the constraint $\del \bcdot \bb{v}=0$)
in a rotating frame. The Coriolis force and the tidal potential
discussed in section~\ref{gov_eq_sec} remain unchanged. This second
approach has recently been adopted by some authors
\citep{umurhan&regev04,lesur&longaretti07}. However, historically,
most authors used the compressible formulation of the MHD equations to
investigate the nonlinear evolution of the MRI. This is
probably due to the rise of computational astrophysical fluid dynamics
that occurred simultaneously to the discovery of the MRI and to the
fact that astrophysical flows are highly compressible in most
instances. Thus versatile codes developed to address astrophysical
fluid dynamics problems are preferentially solving the full set of
compressible MHD equations.

\noindent
There are nevertheless aspects of the problem that are common to both
approaches. One such aspect is the boundary conditions. To understand
them requires to put the shearing box model in its wider context: the
idea is that the entire accretion disk is filled with many such boxes,
with the important point that the statistical properties of the
flow is the same for all boxes. This is valid, for example, in the
case of a turbulent flow whose characteristic scale is smaller than
that of the box. This geometrical picture is illustrated in
figure~\ref{shbc_fig}. The many boxes that constitute the disk are
sliding past one another during a simulation. Thus periodic
boundary conditions are adopted in the azimuthal and vertical
directions. The radial boundary conditions are more complicated, 
though. Consider for example an observer sitting on the inner side (in
the radial, or x, direction) of the box at a azimuthal location. As
time goes on, the observer sees the 
outer radial side of neighboring boxes (located closer to the central
object) sliding past him. Because all these boxes are statistically
identical, the flow it sees in these boxes at any given time is the
same as the flow in {\it its own box, but at the outer radial side of
  its box},
and located at a position in $y$ that varies periodically with time. As we 
see, the radial boundaries conditions are still periodic but in a
peculiar, time varying sense. This is called shearing periodicity. It
has first been introduced in numerical simulations by
\citet{hawleyetal95} and the interested reader is referred to that
paper where the mathematical formulation of the shearing box boundary
conditions is given.

\noindent
Numerical investigations of the nonlinear evolution of the MRI was
quickly undertaken in the early 90's by several
teams. \citet{hawley&stone95} and \citet{hawleyetal95} pioneered the 
way by extending the newly developed algorithm of ZEUS
\citep{stone&norman92a,stone&norman92b} that uses a stable operator
split technique. \citet{brandenburgetal95} used a sixth--order finite
difference scheme that was a precursor to the now well--known {\sc Pencil
Code}\footnote{See
  \url{http://www.nordita.org/software/pencil-code}}. There are now 
several different codes in addition to ZEUS and the {\sc Pencil Code}
that can be used to solve the shearing box equations. Many of them
use finite volume schemes based on the Godunov method
\citep{toro97}. This is the case of ATHENA \citep{stoneetal08}, PLUTO
\citep{mignoneetal07}, 
NIRVANA--III \citep{ziegler04,ziegler08} or RAMSES
\citep{teyssier02,fromangetal06}. Also publicly available is the
pseudo--spectral code SNOOPY\footnote{See 
  \url{http://ipag.osug.fr/~glesur/snoopy.html}} 
that solves the incompressible MHD equations. Such a wealth of
different numerical methods means that published results can now be
carefully tested and are more robust.

\paragraph{Transition to turbulence} 

\begin{figure}
\begin{center}
\includegraphics[scale=0.6]{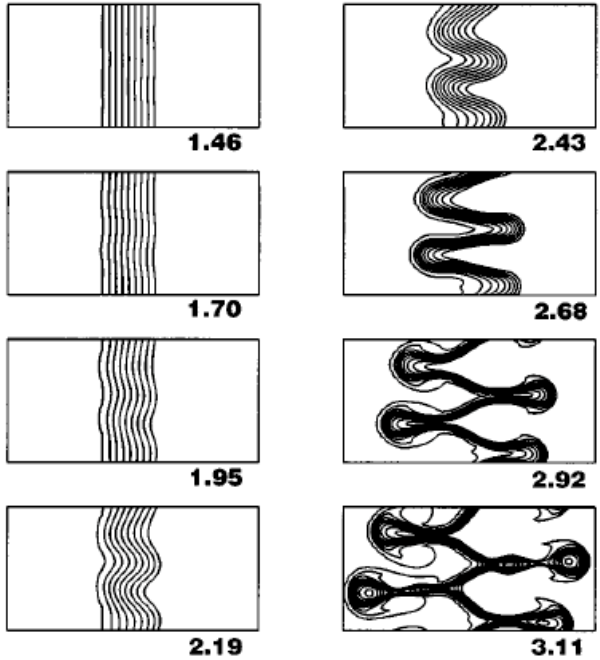}
\caption{Early evolution of poloidal magnetic field lines in an
  accretion disk in the ideal MHD limit. The MRI quickly distorts the
  initially vertical field. Called channel modes, this perturbation
  produces fingers of 
  field and matters that propagate radially. Toward the end of the evolution
  ($t=3.11$), the channel mode amplitude becomes nonlinear and
  vulnerable to parasitic instabilities, as first discussed by
  \citet{goodman&xu94}. Time is labelled in orbits under each
  panels. From \citet{balbus&hawley98}.}
\label{channel_modes_fig}
\end{center}
\end{figure}

\noindent
The first numerical work related
to the MRI appeared as a companion paper to the analytical derivation
of its dispersion relation. \citet{hawley&balbus91} presented 2D
numerical simulations that backed up and extended their linear calculation. In
particular, they confirmed the properties of the channel modes
discussed above. The structure of the magnetic field lines resulting
from the evolution of an initially pure vertical field are illustrated
on figure~\ref{channel_modes_fig}. The oscillating fingers due to the
channel modes create strong magnetic field that moves radially in the
disk, producing large vertical velocity gradients and strong
currents. Both are prone to Kelvin-Helmotz like instabilities, as
investigated in detail by \citet{goodman&xu94}. These ``parasitic''
instabilities grow on top of the regular pattern associated with the
MRI linear modes, disturb the flow and provide a pathway to MHD
turbulence. 

\paragraph{The road to MHD turbulence} The transport properties resulting from
the nonlinear development of 
the MRI can be quantified by measuring the stress tensors discussed in
the introductory chapter of this book. As detailed there, the later is
the sum of the Maxwell and Reynolds stresses. To make connection with
the standard $\alpha$--disk theory presented in
section~\ref{intro_sec}, it is convenient to normalize its value
by the thermal pressure, so that the parameter $\alpha$ can be
measured in numerical simulations according to the relation:
\begin{equation}
\alpha=\frac{\mean{T_{R\phi}^{turb}}}{\mean{P}}=\frac{\mean{-B_RB_{\phi}+\rho
v_R \delta v_{\phi}}}{\mean{P}} \, .
\end{equation}
Early simulations performed in the 90's solidly established some key
results that still hold today. The first were derived using the
homogeneous shearing box threaded by a uniform magnetic field
\citep{hawley&balbus92,hawleyetal95}: 
\begin{enumerate}[topsep=0pt,partopsep=0pt,leftmargin=10pt,itemsep=-2pt]
\item[$\bullet$] When the dissipation coefficients are small enough
  not to affect its linear stage, the nonlinear development of the MRI
  always leads to MHD turbulence that transports angular momentum
  outward. 
\item[$\bullet$] The turbulence is subsonic and the Maxwell stress
  dominates the Reynolds stress by a factor of a few.
\item[$\bullet$] The value of $\alpha$ ranges from $10^{-3}$ to a few
  times $10^{-1}$ depending on the magnetic field strength. $\alpha$
  is an increasing function of the mean magnetic energy.
\item[$\bullet$] The above results hold if the magnetic flux is
  azimuthal instead of vertical, even if $\alpha$ is slightly smaller
  in that case.
\end{enumerate}
These results were obtained in the presence of a nonzero magnetic flux
(either vertical or azimuthal). In the shearing box, such a flux is
conserved during a simulation because of the periodic boundary
conditions. An interesting limit of that case is 
that of a vanishing magnetic flux. This was first considered in the
homogeneous shearing box by \citet{hawleyetal96}. Starting their 
simulations with a random magnetic field, they also found a robust
breakdown of the flow into MHD turbulence. The rate of angular
momentum transport was found to be weaker, though, with $\alpha \sim
10^{-2}$, independent of the field strength. It is important to
realize that this case is qualitatively different from the so--called
net flux case. Indeed, their is no linear instability as such in this
situation. In addition, a dynamo mechanism is needed to sustain the
turbulence over many dynamical timescales. Otherwise, any finite
dissipation will force the flow back to a laminar field as the
magnetic field gradually disapears. The dynamo mechanism operating in
accretion disks was out of the scope of the paper of \citet{hawleyetal96}
but one of their most important finding was that the standard kinematic
dynamo theory was inadequate to describe the field amplification
mechanism: removal of the magnetic feedback on the flow by the Lorentz
force always leads to the decay of the turbulence. 

\noindent
The simulations discussed so far were obtained in the framework of the
homogeneous shearing box, neglecting the vertical stratification of
density. This limitation was soon alleviated in two papers by
\citet{brandenburgetal95} and \citet{stoneetal96} who studied the
development of the MRI in stratified shearing boxes. Both studies
considered an initial magnetic field configuration with vanishing
mean vertical magnetic field. Both found that the MRI leads to
vigorous outward angular momentum transport with typical $\alpha$
values in the range $5 \times 10^{-3}$ to $10^{-2}$. The disk was
found to develop a structure composed of two parts:
\begin{enumerate}[topsep=0pt,partopsep=0pt,leftmargin=10pt,itemsep=-2pt]
\item[$\bullet$] A weakly magnetized layer around the disk midplane
  ($|Z| \leq 2-3 H$), where thermal pressure
dominates over magnetic pressure and with subsonic velocity fluctuations
\item[$\bullet$] Two magnetized layers that form the disk atmosphere
  ($|Z| \geq 2-3 H$) where thermal and magnetic pressure are
  comparable and with sonic velocity fluctuations. 
\end{enumerate}
The case of stratified shearing boxes in the presence of a net
vertical field was considered soon after by \citet{miller&stone00}. It
proved more problematic. As in homogeneous 
boxes threaded by vertical magnetic fields, channel modes were
found to grow to large amplitude. Magnetically dominated regions
appeared and were expelled from the computational box, along with most
of the disk mass, because of magnetic buoyancy. This is in contrast with
homogeneous boxes in which no such vertical gradient is present and
where the channels themselves are destabilized by parasitic
instabilities and turn turbulent. In stratified
boxes, the disk is never able to reach a quasi steady state that
transports angular momentum outward. Because of these difficulties in
finding a gentle turbulent state, this configuration was left aside
from mainstream research for about a decade. This is a pity as this is
a natural magnetic field configuration in PP disks (in which the net
magnetic field is believed to be the remnant of the magnetic field
that was threading the dense core out of which the young star
formed). We shall return to that important problem in
section~\ref{ppdisk_struct_sec}. 

\subsection{Saturation of MHD turbulence: idealized simulations}
\label{saturation_mri_sec}

\begin{figure}
\begin{center}
\includegraphics[scale=0.9]{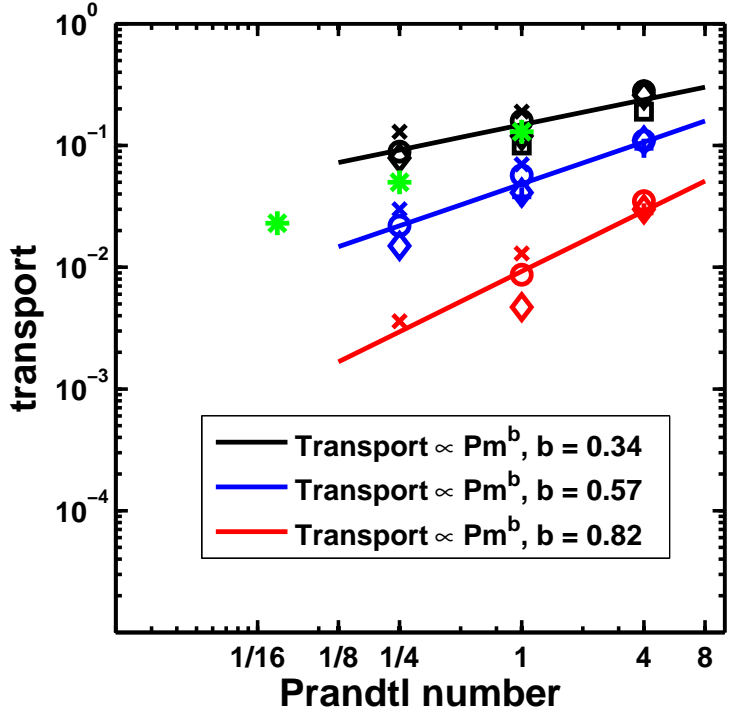}
\caption{Rate of angular momentum transport (as measured using the
  $\alpha$ parameter, see text for details) as a function of the
  magnetic Prandtl number $Pm$ measured using the homogeneous shearing
  box model and in the presence of a vertical magnetic field. Red,
  blue and black symbols correspond to plasma beta $\beta=10^4$,
  $10^3$ and $10^2$, respectively. The different symbol 
  correpond to different Reynolds numbers: $Re=400$ (squares),
  $800$ (plus), $1600$ (diamonds), $3200$ (circles) and $6400$
  (crosses). The green stars are high resolution runs for which
  $Re=c_0H/\nu=20000$ and $\beta=10^3$. Finally, the lines are power
  law fits for each values of $\beta$. From \citet{longaretti&lesur10}.}
\label{longaretti&lesur10_fig}
\end{center}
\end{figure}

\paragraph{MHD turbulence in the ideal MHD limit}
The early simulations described above were done mostly in the limit of
ideal MHD, i.e. neglecting all dissipative terms in the
equations. This was made possible by using numerical schemes that are
stable when solving the Euler fluid equations. As computer power
increased, larger resolution became available and convergence
studies became possible. A problem soon emerged: in a series of papers
devoted to the case of the homogeneous shearing box in the absence of
a net magnetic field,
\citet{fromang&pap07,pessahetal07,simonetal09,guanetal09,bodoetal11}
all found, by 
solving the set of ideal MHD equations, that $\alpha$ was a decreasing
function of resolution. Since the only difference in each sequence of
simulation was the size of the grid cells, these results demonstrated
that the nature of the flow at that small scale affects its large scale
structure. In other words, there is not enough room to separate the
small dissipative scales of the flow from its large and
astrophysically relevant scale. Both scales talk to each
other. This may not be the case in real systems, but was definitely
the case in these simulations. The worrying aspect of the simulations
like that of \citet{fromang&pap07} is that the structure of the flow
at the cells scale is strongly affected by the details of the
numerical scheme. This means in turn that the numerical scheme 
influences the flow largest scale. To solve that problem, a proper
treatment of the small scales is mandatory. This was dealt with by
including physically motivated dissipation coefficients in the
simulations. That small modification to the equations soon lead to new
results as we shall see below.

\noindent
Before we move on, though, we should mention that the convergence
issue described above has now been examined for different field
configurations. It was found that $\alpha$ values are converged in the
presence of a mean toroidal \citep{guanetal09} or vertical
\citep{simonetal09} magnetic field when the resolution is varied. The
reason for the difference with the zero net flux case 
is still unclear but might be due to the presence of
a net flux that constantly help the MRI be reactivated. The
convergence issue of MRI--induced MHD turbulence was also considered
in stratified boxes \citep{davisetal10,shietal10}. Despite having zero
net vertical and toroidal flux, their simulations display nicely
converged $\alpha$ value when the resolution increases. The difference
with unstratified shearing boxes is again not understood, and might
be tied to the existence of a large scale density gradient in the vertical
direction that enables magnetic buoyancy to play a role in the dynamics.

\paragraph{The role of the magnetic Prandtl number}
The numerical convergence issues described above lead to a series of
systematic study of the dependence of the turbulence properties on
small scale dissipation coefficients. As a first step these studies
focused on the effect of a kinematic viscosity $\nu$ and ohmic
resistivity $\eta$. The most robust result that was obtained is that
the rate of angular momentum transport is an increasing function of
the magnetic Prandtl number $Pm$ defined according to the relation
\begin{equation}
Pm=\frac{\nu}{\eta}=\frac{Rm}{Re} \, ,
\end{equation}
where $Rm=c_0H/\eta$ is the magnetic Reynolds number of the flow. This
behavior had been speculated quite early on \citep[see for
example the concluding section of][]{balbus&hawley98} but could not be
studied in the 90's because of limited computational resources. It is
in fact very robust: it was 
observed in the absence of a net magnetic flux \citep{fromangetal07},
in the presence of a net magnetic flux in the azimuthal direction
\citep{simon&hawley09} and in the presence of a net magnetic flux in
the vertical direction
\citep{lesur&longaretti07,longaretti&lesur10}. For illustrative
purposes, figure~\ref{longaretti&lesur10_fig} summarizes the results
of that last study as 
it presents the most extensive parameter space coverage. These first
papers of the $Pm$--effect were devoted to the case of the 
homogeneous shearing box, but \citet{simonetal11a} then showed that
the same trend exists, albeit weaker, when vertical stratification is
taken into account. Their simulations, however, were done for the zero
net flux case. It also covered a limited region of parameter space
because of the large computational cost of such stratified
simulations. Future work is needed.

\paragraph{Ambipolar diffusion and the Hall effect} The magnetic
Prandtl number effect is definitely robust, but it might be irrelevant
in the PP disks context. Indeed, as shown in the previous sections,
the disk locations where ohmic diffusion is the dominant dissipative
term in the induction equation are located well inside the dead zone
where the MRI is stabilized. In turbulent parts of PP disks, the
dominant dissipative term are ambipolar diffusion and the Hall
effect. The effect of ambipolar diffusion has been considered in
details by \citet{hawley&stone98}, \citet{bai&stone11} and
\citet{simonetal13}. Their 
work confirms and extents the results of linear analysis: the maximum
$\alpha$ value is found to drop significantly as $Am$ falls below
unity. In addition, the larger the value of $Am$, the weaker the
magnetic field needs to be for the MRI to operate (with consequently
weaker turbulent transport). 

The case of Hall diffusion is more difficult to handle
numerically. This is because of the presence of a new kind of motion,
the whisler waves, that put strong constraints on the timestep of the
calculation. \citet{sano&stone02a,sano&stone02b} published early
simulations of the MRI in that regime that confirmed the linear
analysis but did not find any significant effect of the Hall term 
during its nonlinear stage. However, as recently pointed out by
\citet{wardle&salmeron12}, the too small Hall term they used in their
simulations makes a definite interpretation of their results quite
difficult. Additional work is needed that focuses on the relevant
regime. 

\begin{figure}
\begin{center}
\includegraphics[scale=0.75]{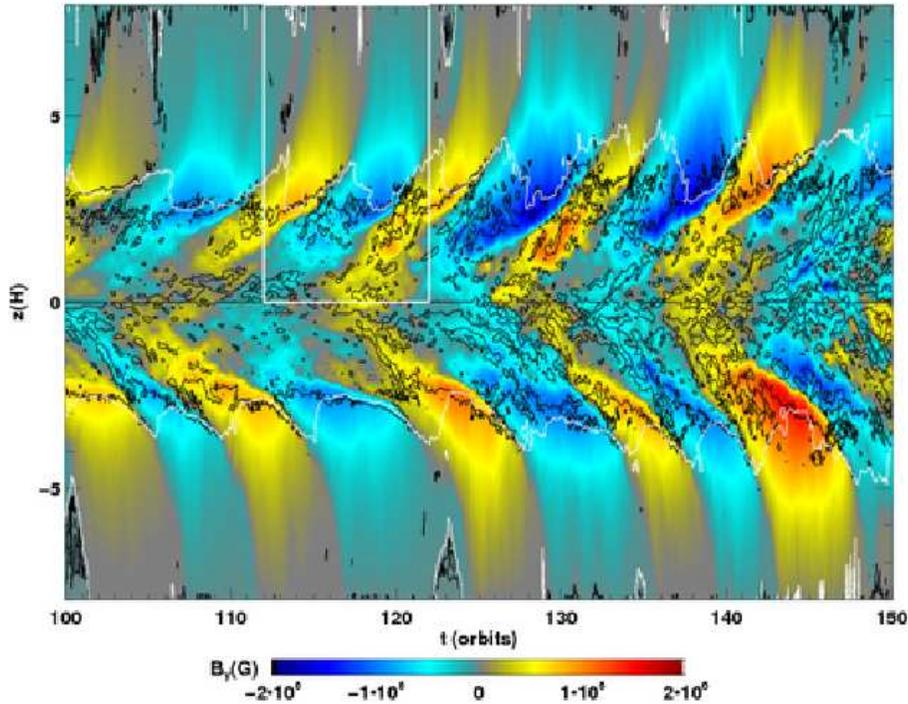}
\caption{Spacetime diagram showing the evolution of the horizontally
  averaged azimuthal magnetic field vertical profile in a typical
  stratified shearing box simulations in the absence of a vertical
  field. A nonzero mean toroidal field is produced in the disk
  midplane before being expelled toward the disk surface layer in a
  quasi periodic way. This typical plot has been named the butterfly
  diagram and is believed to be a manifestation of some sort of mean
  field dynamo. From \citet{shietal10}.}
\label{butterfly_fig}
\end{center}
\end{figure}

\paragraph{Current status, open questions} While some results are now firmly
established, it is fair to say that we still lack a complete
understanding of the saturation of MRI--induced MHD turbulence as a
function of the disk parameters. Below is a list of some of the most
critical questions and, when possible, the immediate prospect for
their solutions:
\begin{enumerate}[topsep=0pt,partopsep=0pt,leftmargin=10pt,itemsep=-2pt]
\item[$\bullet$] The sensitivity of angular momentum transport to $Pm$
  immediately raises the question of the asymptotic behavior of the
  flow in the small $Pm$ limit that is relevant for PP
  disks\footnote{We have seen in section~\ref{intro_sec} and
    \ref{mri_lin_sec} that the electron fraction can be 
    extremely small. As a result, the resistivity is much larger than
    the viscosity in PP disks and the magnetic Prandtl number is much
    smaller than unity}. The best
  simulations are currently addressing that question, but some hints
  are already present in the literature: for example, in the case of a
  pure toroidal magnetic field, figure 7 in \citet{simon&hawley09}
  shows using blue symbols a sequence of $\alpha$ values computed for
  $Rm=3200$ and decreasing values of $Pm$ (or, equivalently,
  increasing $Re$). Clearly, it is tempting to say that $\alpha
  \rightarrow 0.02$ when $Pm \rightarrow 0$. Similarly, in the case of
  a vertical magnetic field, figure~\ref{longaretti&lesur10_fig} shows
  early suggestions of convergence: the series of runs having $Rm=1600$ and
  $\beta=10^3$ (blue and green colors) appears to converge toward
  $\alpha=2 \times 
  10^{-2}$ at low $Pm$. Whether this trend is confirmed at much lower
  $Pm$ and its dependence on the value of $Rm$ should be the focus of
  future work, but current results indicate that transport at low $Pm$
  is possible in the presence of a net flux. 
\item[$\bullet$] In the absence of a net flux, the $Pm$ effect is
  spectacular: when $Pm<1$, the turbulence dies off and the flow
  return to its laminar state \citep{fromangetal07}. This is not
  without similarities with the known results that small scale dynamo
  in incompressible fluids is more difficult to trigger
  \citep{schekochihinetal04} at low $Pm$: in that limit, there exists
  a critical magnetic Reynolds number $Rm_c$ above which small scale
  dynamo is observed to grow that increases as $Pm$ decreases
  \citep{schekochihin07a}. Is it 
  the same for the MRI? What are the properties of the dynamo mechanism
  that operates in this regime? The computational requirements to
  address that problem are massive, and it is not clear that brute
  force is the answer. Recently, \citet{heraultetal11} proposed a
  novel approach to the problem and found that dynamo cycles 
  exist even when dissipation coefficients are large. Whether the
  properties of such cycles are imprinted into fully developed MHD
  turbulence remains to be demonstrated, but the results appear
  promising at this stage. 
\item[$\bullet$] The influence of density stratification on these
  results is a third important open issue. Several papers reported
  the appearance of a mean toroidal field, the strength of which is
  modulated in time
  \citep{brandenburgetal95,gressel10,shietal10,davisetal10,simonetal12}. This 
  mean magnetic field is expelled toward the disk upper 
  layers producing a regular pattern in the spacetime diagram (see
  figure~\ref{butterfly_fig}). This properties of the flow has been
  named the butterfly diagram. The reason for its appearance is not
  firmly established and its potential role in the dynamo mechanism is
  still debated.
\end{enumerate}

\subsection{Global disk simulations of turbulent PP disks}

\begin{figure}
\begin{center}
\includegraphics[scale=0.1]{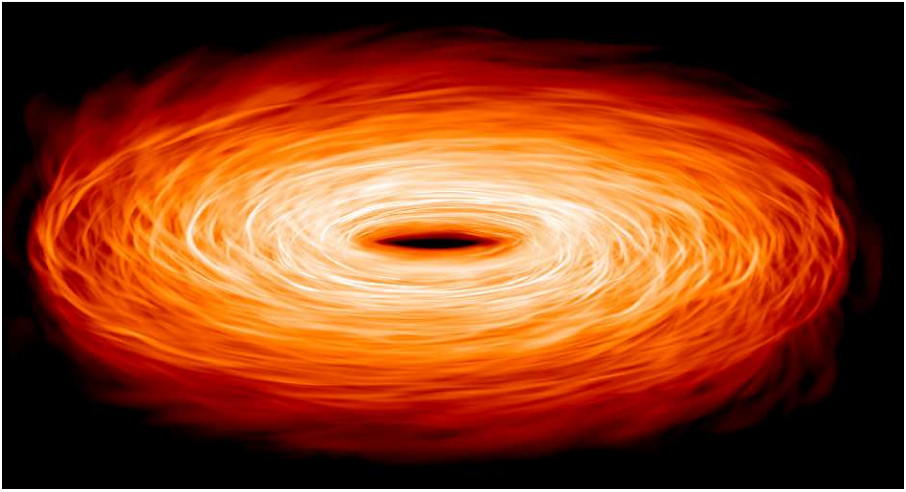}
\caption{Snapshot of the magnetic pressure produced in a typical MHD
  global simulation of a turbulent PP disk. Simulations data have been
  post--processed using the code RADMC
  (\protect\url{http://www.ita.uni-heidelberg.de/~dullemond/software/radmc-3d/})
  to produce the 3D rendering. Image courtesy Mario Flock.} 
\label{mario_disk_fig}
\end{center}
\end{figure}

So far the discussion has been restricted to local studies, both 
analytically in our derivation of the stability criterion as well as
numerically in our summary of the most salient results of shearing
boxes numerical simulations. However, PP disks are large scale objects
in which the key parameters (surface density, temperature, ionization
fraction, magnetic flux) potentially evolve with position and
time. However useful a local approach is to our understanding of the
turbulence properties, we ultimately want to compare these models with 
observations of PP disks as a whole. This is why several teams
developed global numerical simulations of turbulent accretion disks.

\paragraph{The issue of resolution} The rule of thumb when discussing
global simulations of PP disks is that the numerical resolution of the
calculation is always too low! Ideally, one would want to make a model
of a disk covering the planet formation region, say from $1$ to $10$
AU of the central star and covering a few vertical scaleheights on
both sides of the disk midplane (say, $\pm5$ H to fix ideas). Typical
PP disks have disk aspects ratio $H/R$ that range from about $0.05$ to
$0.1$. For the thicker value (which is easier to handle
numerically), this means the simulations would cover about $10$
scaleheights in the vertical direction and $25$ scaleheights in the
radial direction. Modern day numerical simulations have a resolution
of a few hundreds cells in each direction. This means that the most
resolved global simulations today can at best achieve a 
resolution of $10$ to $20$ cells per scaleheight. This should be
contrasted with the resolution of about $128$ to $256$ cells per
scaleheight used in local simulations to study the saturation
properties of the turbulence as a function of microscopic dissipation 
coefficients. Clearly, it is impossible at present time to
properly include dissipation in global simulations and be in a
parameter regime relevant for PP disks. One should instead rely on
numerical dissipation to stabilize the numerical scheme when
performing such simulations, and hope for the best... 

\paragraph{Typical models and main results} Due to the limitations
detailed above, global simulations of turbulent PP disks still face
drastic limitations even when resolution is reduced to its minimum
acceptable values. Simplifications have to be made. One of them is
an oversimplified thermodynamics. Since PP disks are mostly passive
disks heated by their harboring star, a locally isothermal equation of
state is often used. For simplicity, the gas sound speed obeys a power
law  
\begin{equation}
c_s^2(R)=c_0^2 \left( \frac{R}{R_0} \right)^{q} 
\label{cs_def}
\end{equation}
in most published simulations. $q=-1$ is often adopted. Another
limitation is provided by the limited radial extent of the
simulations. Published simulations usually 
cover a radial extent of about a decade. This means that radial
boundaries (the treatment of which is not without problem but is beyond the
scope of this lecture) are never far from the bulk of the disk. They
might influence the flow properties. This is a potential problem that
should be kept in mind. Another difficulty with global simulations is
their large range in dynamical timescales: the orbital time is much
larger at the outer edge than at the inner edge. This means that the
simulations requires to be evolved for many orbital period at the
inner edge before reaching a quasi steady state at the outer
edge. This is not without consequences in terms of computing resources
needed for such simulations.

\noindent
Despite these difficulties and caveats, papers dedicated to
studying the structure of the flow in turbulent PP disks by means of
global simulations have been published by several authors. The first
simulations were performed in the cylindrical limit
\citep{steinacker&pap02,pap&nelson03a,wintersetal03,sorathiaetal12}. They
were built as an extension of the first pioneering simulations of
\citet{armitage98} and \citet{hawley01} that were more specifically
aimed at studying black hole accretion. Recent work also includes vertical
stratification
\citep{fromang&nelson06,sorathiaetal10,beckwithetal11,flocketal11,fromangetal11}. An
example of the results of such global simulations is given in
figure~\ref{mario_disk_fig}. The main results are:
\begin{enumerate}[topsep=0pt,partopsep=0pt,leftmargin=10pt,itemsep=-2pt]
\item[$\bullet$] Typical $\alpha$ values are in the range $5 \times
10^{-3}$ to a few $10^{-2}$, consistent with shearing boxes
simulations performed at the same resolution and with the same
B--field configuration.
\item[$\bullet$] The results appears to display {\it numerical}
  convergence for resolution of $32$ cells per $H$ or higher
  \citep{sorathiaetal12}. 
\item[$\bullet$] There is a correlation between the local stress
tensor and the local vertical flux of magnetic field
\citep{sorathiaetal10, beckwithetal11}
\item[$\bullet$] There are significant turbulence fluctuations at
scales larger than the typical disk scaleheight
\citep{beckwithetal11,flocketal12}. As a result, a minimum azimuthal 
size of the computational box of $\sim \pi/2$ is required.
\item[$\bullet$] The existence of meridional circulation seems to be
excluded in fully turbulent disks \citep{flocketal11,fromangetal11},
which is a significant departure from the expectations of viscous disk
theory.
\end{enumerate}
Before closing this section, however, it is important to stress that
most global simulations published so far have assumed ideal MHD, with
effectively zero net vertical flux. This is a significant limitation
compared to the wide range of configurations that have been probed
using shearing boxes simulations. Much work remains to be done to
improve the quality of global simulations of PP disks. 

\section{Consequences for protoplanetary disks and planet formation}

Sections~\ref{mri_lin_sec} and \ref{mhd_turb_sec} summarize the
understanding we have gained during the last couple of decades about 
MRI--driven angular momentum transport in PP disks. The situations
considered in these sections are rather academic: idealized situations
are constructed with the aims to understand the fundamental
properties of the MRI and its nonlinear consequences. The next stage
is of course to build realistic models of PP disks based on these
findings. Such models can then be used to refine our understanding of
planet formation scenarii. The purpose of the present section is to
review these topics and to highlight some of the outstanding problems
that are still to be solved. 

\subsection{Protoplanetary disks structure}
\label{ppdisk_struct_sec}

\paragraph{Layered accretion} Since the seminal paper of
\citet{gammie96}, layered accretion is the classical paradigm
describing the flow in PP disks: it consists in a fully turbulent
inner disk, while at larger radii, the flow is laminar in the bulk of
the disk and turbulent in its surface layers. Local simulations have
investigated the z-dependence of the flow structure at radii where a
dead zone is present while global simulations have focused on the
dynamics taking place at the inner edge of the dead zone. The main
results of these simulations are: 
\begin{enumerate}[topsep=0pt,partopsep=0pt,leftmargin=10pt,itemsep=-2pt]
\item[$\bullet$] Turbulent motions in the active surface layers of the
  disk excite waves that propagates deep into the dead zones
  \citep{fleming&stone03}. These waves create a sustained angular
  momentum transport due to the associated Reynolds stress. Typical
  $\alpha$ values that can be attributed to such waves are of the
  order of $10^{-5}$ to $10^{-6}$ and depend on the dead zone mass.
\item[$\bullet$] Most of the published simulations are performed using
  an isothermal equation of state. This is not appropriate for the
  optically thick inner parts of the disk, where cooling is slow and
  heating is due to the dissipation of turbulent energy. Some authors
  \citep{hirose&turner11,flaigetal10,flaigetal12} recently relaxed that
  approximation in stratified shearing boxes calculations. They
  reports quantitative changes only, the qualitative picture of
  layered accretion being unaltered. However, recent mean field models
  of the coupling between dynamics and thermal processes suggest that
  new effects might affect the global structure of the disk
  \citep{latter&balbus12}. The confirmation of these results requires 
  global simulations. 
\item[$\bullet$] In agreement with the expectations resulting from 2D
  numerical simulations of viscous disks \citep{varniere&tagger06},
  pressure maxima are found to develop at the dead zone inner edge
  \citep{nataliaetal10}. Using similar simulations,
  \citet{lyra&maclow12} found that vortices grow at that location as a
  result of the Rossby wave instability \citep{lovelaceetal99}. Such
  vortices are efficient trapping structures of dust particles
  \citep{barge&sommeria95} and could thus be ideal locations to initiate
  planet formation.
\end{enumerate}
It should be noted, though, that the published simulations lack
two key physical ingredients that are potentially important. One is
the absence of a mean vertical field, which is unfortunate as this is
the most natural magnetic field configuration. The second is that
magnetic diffusivity is treated as being due to ohmic diffusion
only. Ambipolar diffusion and the Hall effect, which we have seen are
important in the disk upper layers, are simply ignored. Including both
ingredients is only starting and we briefly mention the exciting early
results that have been obtained in the next section.

\paragraph{Disk with vertical magnetic fields: linking disk and
jets/winds}

\begin{figure}
\begin{center}
\includegraphics[scale=0.35]{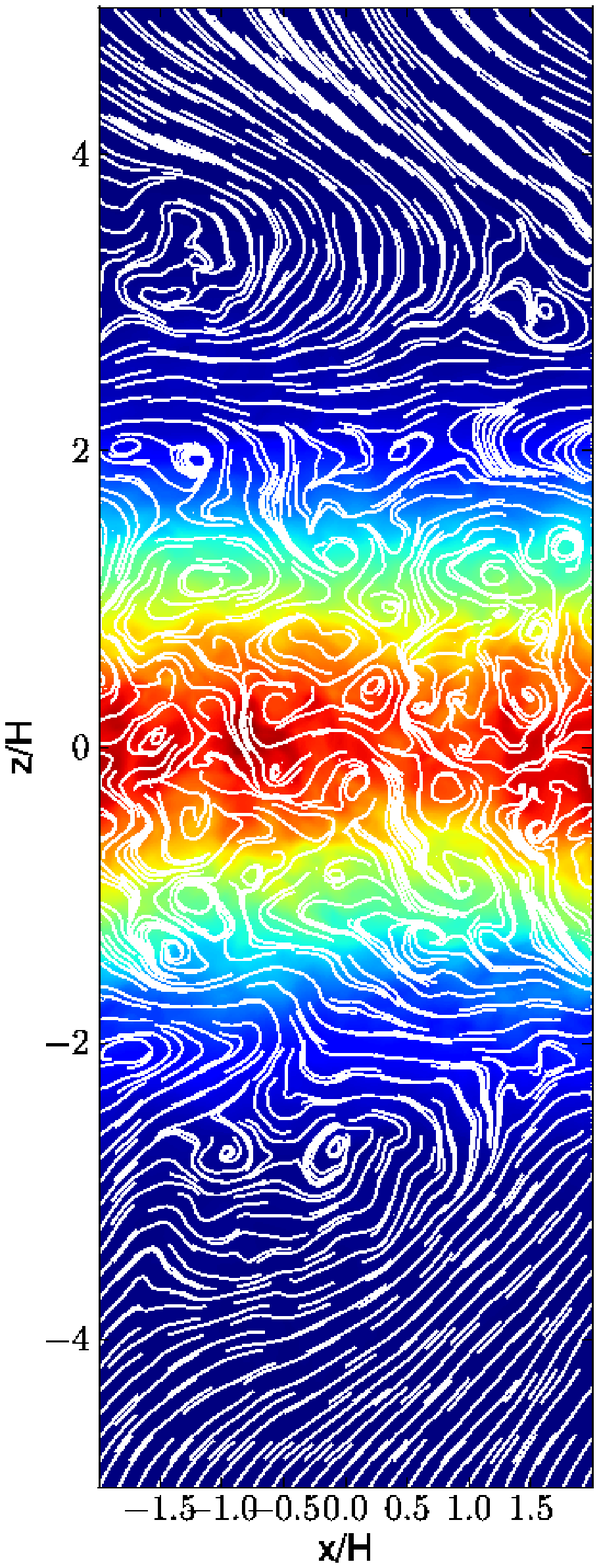}
\includegraphics[scale=0.35]{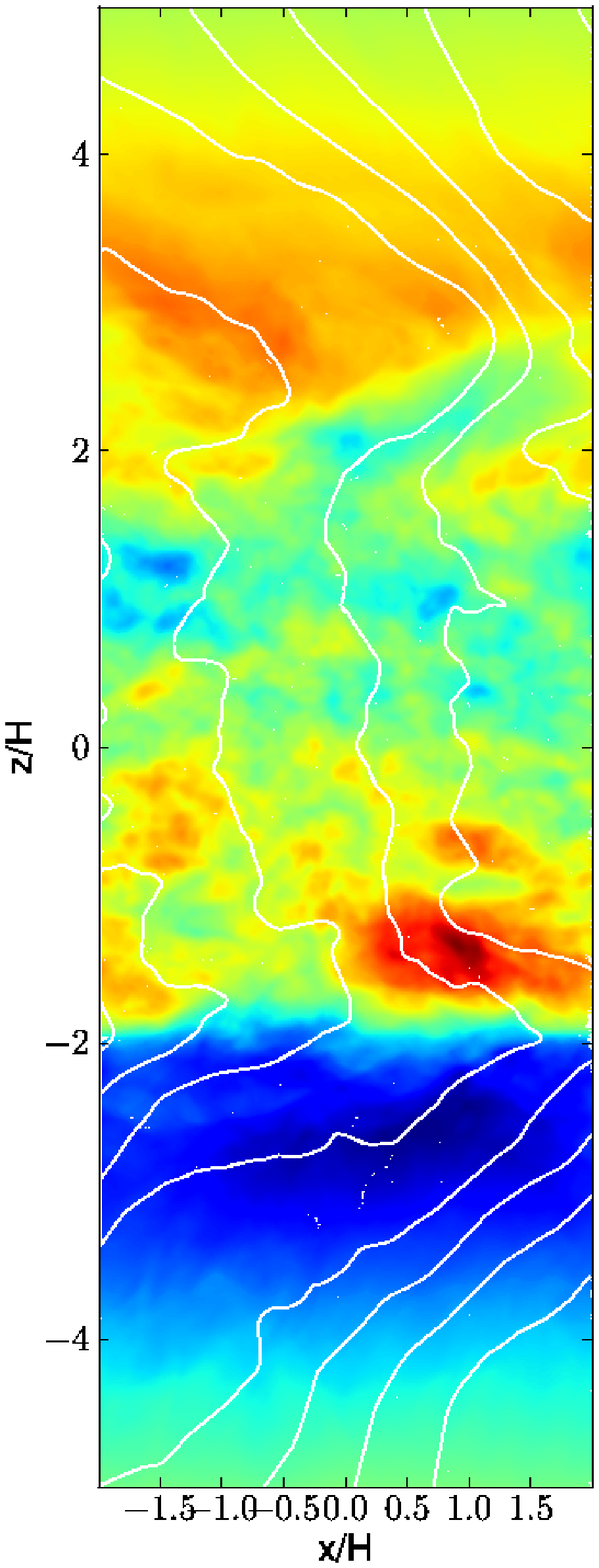}
\caption{Structure of the horizontally averaged poloidal streamlines
  ({\it left panel}) and poloidal magnetic field lines ({\it right
    panel}) obtained in a stratified shearing box simulation in the
  presence of a net vertical field (such that the midplane
  $\beta=10^4$) overplotted on the density distribution ({\it left 
    panel}) and azimuthal magnetic field distribution ({\it right
    panel}). From \citet{fromangetal13}.}
\label{lines_burst_fig}
\end{center}
\end{figure}

As discussed in section~\ref{turb_history_sec}, the most
natural configuration of a disk threaded by a nonzero vertical
magnetic field was long put aside due to early numerical
difficulties. In addition to being important for our overall
understanding of angular momentum transport in disk, such
configurations are also known to lead to jets/winds launching by
the disks (see the chapter by J.Ferreira for more details). Such models
generally require two ingredients: a near equipartition magnetic 
field ($\beta \sim 1$) as well as efficient magnetic field
radial diffusion. The latter is required to prevent the magnetic field
lines from accumulating near the central star. It is believed to
result from disk turbulence. The problem is that the MRI is a weak
field instability that is stabilized when the field reaches
equipartition (see section~\ref{mri_lin_sec}). Thus the dynamical
state of the disk in that regime is subject to large
uncertainty. Recently, a set of papers returned to that particular
geometry using stratified shearing boxes simulations (still with a
simplified isothermal equation of state). Several results were obtained:
\begin{enumerate}[topsep=0pt,partopsep=0pt,leftmargin=10pt,itemsep=-2pt]
\item[$\bullet$] The difficulties encountered by
\citet{miller&stone00} can be solved by letting the flow develop a
quasi steady state in the absence of vertical field, before gradually
adding that field over a few dynamical timescales
\citep{bai&stone13a}. This procedure prevents the MRI channel flows from
growing and destroying the disk. Vigorous turbulence ensues when the
disk reaches a quasi steady  state, with $\alpha$ values up to $1$
when the midplane magnetic field strength is such that $\beta=10^2$.
\item[$\bullet$] Regardless of the field strength, the disk launches a
powerful outflow out of the computational box
\citep{suzuki&inutsuka09,suzukietal10,moll12,bai&stone13a,fromangetal13,lesuretal13},
the properties of which resembles the classical disk wind solution
first described by \citet{blandford&payne82}. The time averaged flow
structure obtained in a typical simulation of that kind is illustrated on
figure~\ref{lines_burst_fig}. The  mass loss rate is significant but
has been found to be sensitive to 
numerical details of the simulations such as box size. Whether the
outflow is successful or fall back on the disk also remains to be
established. Both aspects might be related to limits of the shearing
box model and will probably require global simulations to be
solved. Needless to say, such simulations represent a formidable task
at the limit of present day supercomputers.
\end{enumerate}
These recent results represent significant improvements in our ability
to design realistic numerical simulations of turbulent accretion
disks. They are not good representations, however, of the planet
forming region in PP disks. As discussed extensively above, non-ideal
MHD effects are dynamically important: ohmic diffusion dominates in
the disk midplane while the Hall effect and ambipolar diffusion are
important in the disk upper layers. Recently, \citet{bai&stone13b}
considered such a situation by performing simulations of stratified
shearing boxes with a weak net vertical field ($\beta=10^5$ in the
disk midplane) including ohmic and ambipolar term in the induction
equation. The vertical profile of both terms were interpolated from a
pre--computed lookup table which was calculated from evolving a large
chemical network to equilibrium. They find that ambipolar diffusion
makes qualitative changes to the structure of the flow: the otherwise
turbulent surface layers of PP disks is stabilized by the ambipolar
term and an outflow that takes away mass and angular momentum is
launched from these laminar layers. The disk 
evolves toward a steady state with efficient outflow mediated angular
momentum transport. This is a major evolution compared to the
standard dead zone paradigm pictured in figure~\ref{flow_prop_fig}. Its
robustness should be carefully assessed by future work. This will
certainly require combining shearing boxes and global numerical
simulations and represents a major challenge for future years.

\paragraph{Other transport mechanisms}
\label{hydro_transport_sec}

So far, we have concentrated on the MRI and its consequences as the
only route to extract angular momentum from PP disks. What about pure
hydrodynamical processes? Such processes, if they exist, could be
dominant in the dead zone where the MRI does not operate. In other
disk regions, they will coexist with MRI--induced turbulence and we
may ask if such an interaction has important dynamical consequences.

\noindent
The first question to address is that of the existence of such
hydrodynamical processes. As noted in the introduction and in
section~\ref{mri_lin_sec}, accretion disks are 
linearly stable according to the Rayleigh criterion. However, because
of the enormous value of the Reynolds number, they might be
nonlinearly unstable (as simple Couette flow in the laboratory). This
has been the subject of intense controversy in the last 
twenty years. On the theoretical side, the first results have
suggested that the flow remains laminar \citep{hawleyetal99} because
of the stabilizing effect of the Coriolis force. This has recently
been confirmed both numerically \citep{lesur&longaretti05} as well
as using semi-analytical techniques \citep{rinconetal07}. Experiments
of Taylor--Couette flows confined between two rotating cylinders have
also recently been designed to study that problem. They have yielded
conflicting results: \citet{jietal06} and \citet{schartmanetal12} find
the flow remains 
laminar up to Reynolds numbers of order a million while
\citet{paoletti&lathrop11} report signatures of a turbulent flow for
similar Reynolds numbers. The interpretation of these experiments,
however, is complex because of the presence of axial boundaries that
produce Ekman layers \citep{avila12}. The nature of 
the flow in these Taylor--Couette experiment is thus unclear. Progress
should be made on this issue before applying the results to PP disks. 

\noindent
When the equation of state of the flow is not barotropic, vorticity can
be produced in the flow. This opens up the possibility for the
growth of the baroclinic instability
\citep{petersenetal07,lesur&pap10}. Detailed studies of the
saturation of the baroclinic instability are required, 
but the first results \citep{raettigetal13} suggest $\alpha$
values of order a few times $10^{-3}$. In any case, the dynamical
influence of the baroclinic instability will be confined to the dead
zone, as it is completely overwhelmed by MRI--induced MHD turbulence
in the regions where the latter operates \citep{lyra&klahr11}.

\noindent
When their mass becomes large, PP disks start to be influenced by
their own gravitational field. The importance of self--gravity is
quantified by a dimensionless number called the Toomre parameter: 
\begin{equation}
Q=\frac{c_0 \kappa}{\pi G \Sigma} \, .
\end{equation}
The rule of thumb is that self--gravity is negligible when $Q \gg
1$. When it approaches unity, gravitational instabilities
develop. They take the form of spiral arms as seen in galaxies. In
order to get a feel for when a given disk enters such a regime, it is
enlightening to express $Q$ in terms of both the central and the disk
masses. Using the explicit radius dependency for $\Omega$ and the
relation $H=c_0/\Omega$, we can write 
\begin{equation}
Q \sim \left( \frac{H}{R} \right)\left( \frac{M}{M_d} \right) \, ,
\end{equation}
where $M_d=\pi R^2\Sigma$ is the approximate disk mass contained in the
disk within a radius $R$. For PP disks where $H/R \sim 0.1$, the above
expression 
states that self--gravity effects become important when the disk mass
becomes of order one tenth of the central mass or larger. This is in
general larger than the typical values of PP disks masses, although
the latter are still uncertain and subject to controversy
\citep{berginetal13}. At the limit when $Q$ reaches one or smaller
values, the disk is expected to become strongly unstable to
non--axisymmetric modes and fragment. This has led to speculations
that planets might form as a result of gravitational
instabilities. The discussion of whether this actually happens or not
in real systems is much beyond the scope
of this lecture and could in fact be a lecture by itself. The
interested reader might want to consult the short review of that
problem by \citet{stamatellos13} and references therein. For the
purpose of this course, we just note that the issue of planet
formation is far from being settled, as shown by two recent papers
\citep{paardekooper12,meru&bate12} that illustrate some of the
numerical difficulties that still remains despite years of
investigations. 

\noindent
This discussion on the possible links between planet formation and
gravitational instability naturally leads us to the examine the
possible effect the MRI might have on planet formation processes. This
is the purpose of the following section.

\subsection{Planet formation}

According to the recent results of the Kepler mission
\citep{batalhaetal13}, exoplanets are ubiquitous in the
universe. While the exact formation mode is still debated, there is a 
widespread consensus that planets are born in PP disks before gas
dissipates. The two competing (and maybe not mutually excluding)
scenarii of planet formation rely on gravitational instability
(see section~\ref{hydro_transport_sec} above) and on the 
so--called core accretion model \citep{pollack96}. In the latter
scenario planets gradually increase in mass by accumulation of solids,
forming 
planetesimals, earth mass planets, and, when reaching a large enough
mass, accreting an envelope to form gas giant planets like
Jupiter. The purpose of the present section is to illustrate the
typical consequences of the MRI on that process through two examples:
dust dynamics (and more specifically, dust settling toward the disk
equatorial plane) and planet disk interaction.

\noindent
There are many more aspects of the problem that are not covered
here. Particularly important are the formation of planetesimals
\citep{johansenetal07} and their subsequent evolution \citep[see for
  example][and references therein]{gresseletal12}, for which the
influence of MHD turbulence is important. The MRI has
also been considered in the context of on many other aspects of planet
formation, such as collisions between dust particles
\citep{carballidoetal10} or chondrules formation \citep{mcnallyetal13}
to name just a few examples. 

\paragraph{Dust dynamics in turbulent PP disks}

\noindent
Dust dynamics is dominated by three effects: vertical settling
toward the disk midplane, radial migration toward the central star and
growth through coagulation. Here we focus on the first two processes,
which are due to the drag force between gas and solid particles. For
illustrative purposes, we consider the case of small dust
particles only. When their size is smaller than the molecules mean free
path (see section~\ref{intro_sec}), we are in the so--called Epstein
regime, in which the particles feel a drag force $\bb{F_d}$ from the
gas that takes the simple form
\begin{equation}
\bb{F_d}=-\frac{\bb{v}-\bb{v_d}}{\tau_s} , \,\,\, \textrm{where} \,\,\, \tau_s=\frac{\rho_s a}{\rho c_s} \, .
\end{equation}
In the above expression, $\bb{v_d}$ stands for the dust velocity and
$\tau_s$ is the dust stopping time. This is the typical time it takes
for dust particles initially at rest to reach the local gas
velocity. It depends on the dust particle internal mass density
$\rho_s$ and their size $a$. $\tau_s$ can be compared with the local
dynamical 
time by defining the Stokes number $St=\Omega \tau_s$. When $St \ll
1$, the stopping time is much smaller than the orbital period
$T_{orb}$ and the dust essentially follows the gas. When $St \sim 1$
or larger, $\tau_s$ becomes comparable to $T_{orb}$ and dust and gas
start to decouple.

\noindent
The consequences of that friction force are simple to describe. Let's
consider a single particle initially at rest (in a rotating frame) at
a given radius $R_d$ and height $Z_d$ above the midplane. In the
absence of gas, this particle rotates around the central star on
inclined orbits at the Keplerian frequency. In a frame rotating at
that frequency, the particle oscillates around the disk midplane. In
the presence of gas in hydrostatic equilibrium, two effects appear:
the moving particle experience a head wind due to the gas as a result
of this oscillation. This force damps the particle oscillation. As a
result, the particle falls toward the disk midplane, a process known as
gravitational settling. A second effect arises 
because gas rotates at sub-Keplerian frequencies around the central
object (this is due to partial support due to the negative radial pressure
gradient). Thus dust particles rotate faster than the gas, experience
a head wind and lose angular momentum, migrating inward in the disk.

\begin{figure}
\begin{center}
\includegraphics[scale=0.065]{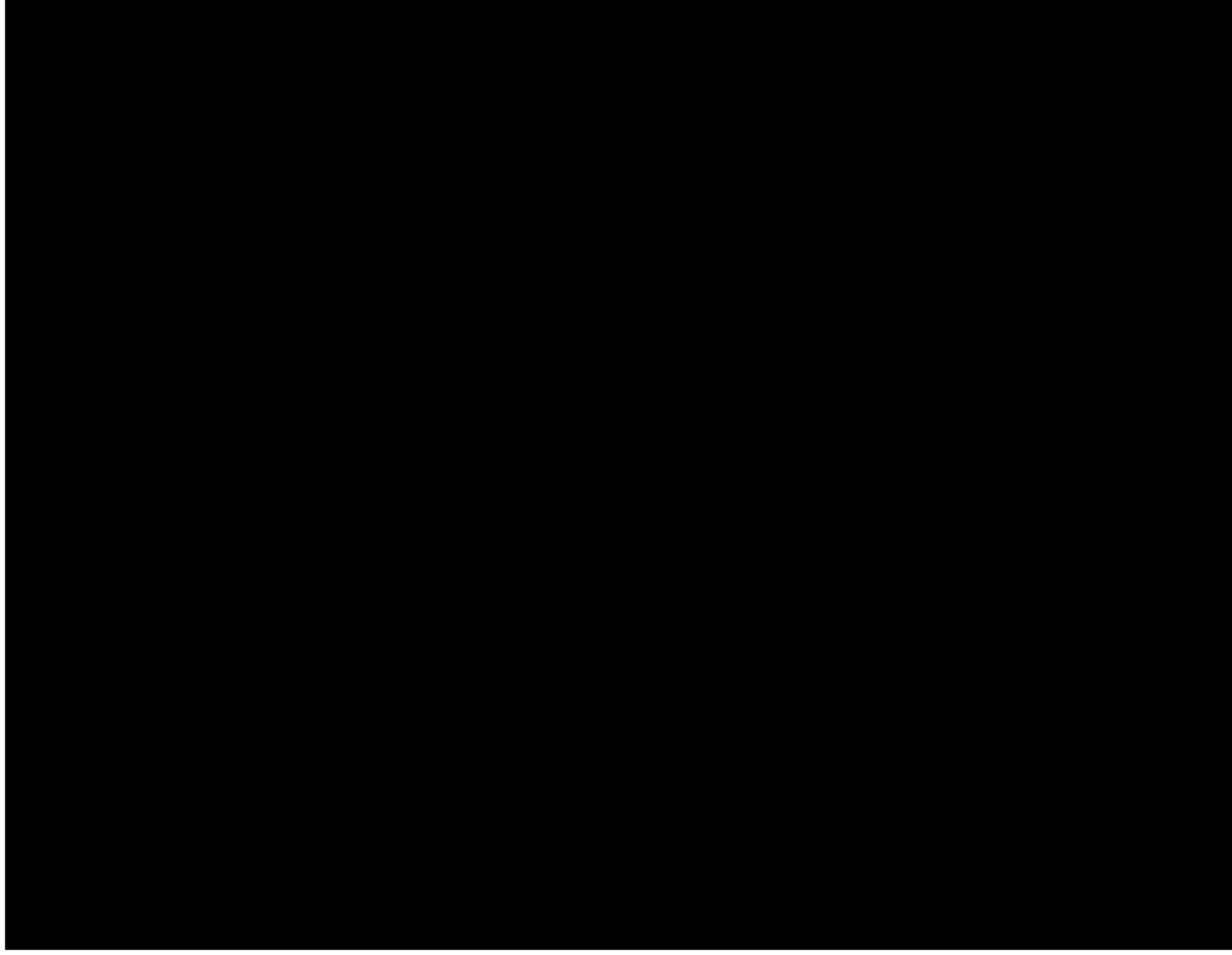}
\includegraphics[scale=0.065]{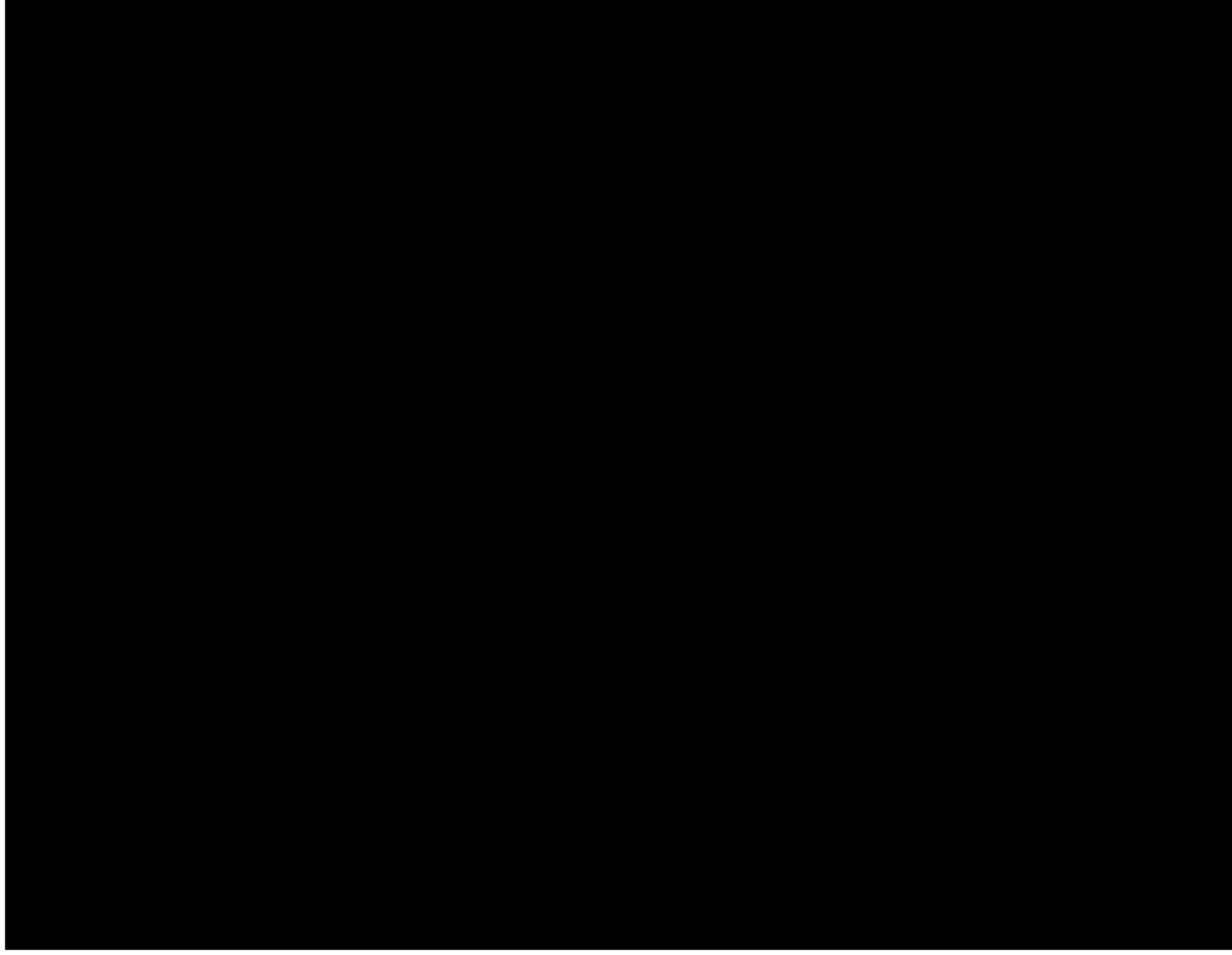}
\includegraphics[scale=0.065]{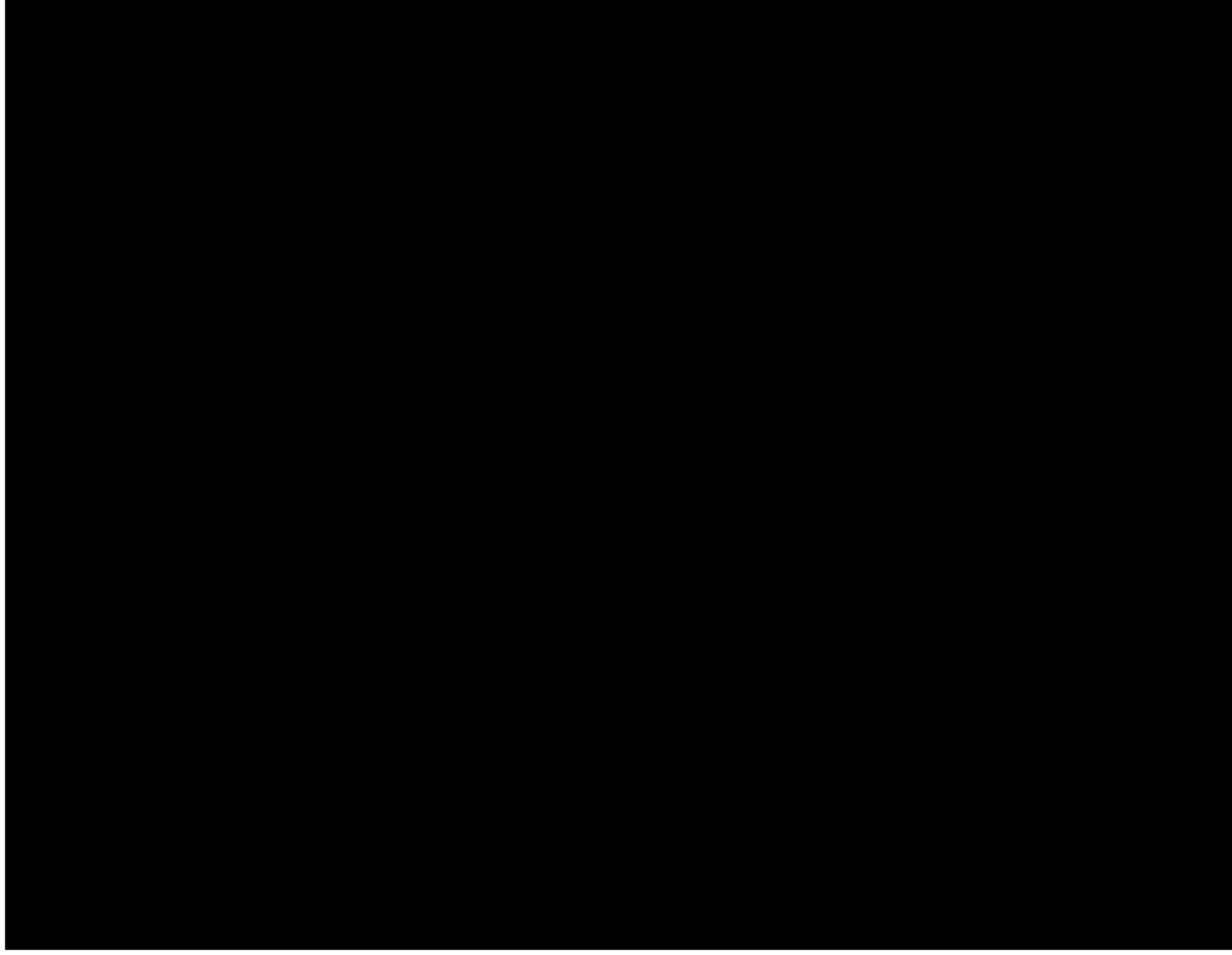}
\caption{Dust distribution in the $(R,Z)$ plane in a stratified
  shearing box numerical simulations. From left to right, the
  different panels correspond to $St=0.001$, $0.01$ and $0.1$. From 
\citet{fromang&pap06}}. 
\label{larger_snapshot}
\end{center}
\end{figure}

\noindent
Turbulence affects the above picture by randomly diffusing
particles. In the case of gravitational settling, it prevents
dust particles from collapsing into an infinitely thin layer. The 
effect of turbulence can be modeled as a diffusive process. If $D$ 
stands for a turbulent diffusion coefficient, the vertical evolution
of the dust density can be described by the following 
partial differential equation
\citep{dubrulleetal95,schapler&henning04,dullemond&dominik04}:
\begin{equation}
\frac{\partial \rho_d}{\partial t}-\frac{\partial}{\partial z}(z
\Omega^2 \tau_s \rho_d)= \frac{\partial}{\partial z} \left[ D \rho 
  \frac{\partial}{\partial z} \left( \frac{\rho_d}{\rho} \right)
  \right] \, ,
\label{diff_advec_eq}
\end{equation}
where $\rho_d$ is the dust particle density. This equation models the
balance between vertical settling and turbulent diffusion. In steady
state, assuming that the diffusion coefficient $D$ is constant and the
gas density vertical profile is a Gaussian, Eq.~(\ref{diff_advec_eq})
can be integrated analytically to give
\begin{equation}
\rho_d=\rho_{d}^0
\exp \left[- S \left(
  \exp\left(\frac{Z^2}{2H^2}\right)-1\right) -\frac{Z^2}{2H^2} \right]
\propto \exp \left[-(1+S)\frac{Z^2}{2H^2}\right]
\label{dust_prof_eq}
\end{equation}
where $\rho_{d}^0$ is the midplane dust density. The parameter
$S=St^0/(D/c_sH)$ measures the relative importance of the turbulent
diffusion and the midplane Stokes number \citep{jacquetetal12}. The
last part of Eq.~(\ref{dust_prof_eq}) is obtained as a result of an
expansion in the strong settling limit $Z \ll
H$. Eq.~(\ref{dust_prof_eq}) quantifies the effect of turbulence on
gravitational settling: when $S \gg 1$, the dust layer has a thickness
$H_d=H/\sqrt{1+S}$, i.e. it is much smaller than that of the gas
\citep[one should be careful, though, as the fluid approximation
used to establish eq.~(\ref{diff_advec_eq}) breaks down in the large
Stokes number limit, see e.g.][]{garaudetal04,youdin&lithwick07}. When
$S \ll 1$, the assumption leading to the asymptotic expansion breaks
down and the vertical profile for the dust is more complex but the
qualitative result that $H_d \sim H$ still remains. The simple
arguments provided here for illustrative purposes can be made more
rigorous and can be extended to larger particles and to the effect of
turbulence on radial dust migration \citep{youdin&lithwick07}. In all 
cases, turbulence broadens the spatial distribution of a population of
grains initially located at a single position in the disk. 

\noindent
Validation of these ideas requires detailed MHD numerical simulations in
which the turbulent flow structure is self-consistently
calculated. Such simulations devoted to studying vertical settling 
have been performed by a number of groups using local simulations 
in the shearing box
\citep{johansen&klahr05,fromang&pap06,turneretal06,turneretal07,turneretal10} 
as well as global simulations of fully turbulent disks
\citep{fromang&nelson09}. The effect of turbulence on radial migration
has been analyzed by \citet{carballidoetal11}. The conclusion of all
these studies is the same: the dynamics induced 
on dust particles by the turbulent flow is well described by a
diffusive process. In most instances, the simulations demonstrated
quantitative agreement with analytical theories, such as that of
\citet{youdin&lithwick07}.

\noindent
These prescriptions are now being incorporated into more realistic,
but semi-analytical models of PP disks that include dust coagulation 
\citep{birnstieletal12,charnoz&taillifet12} and permit long
integration times. 

\begin{figure}
\begin{center}
\includegraphics[scale=0.35]{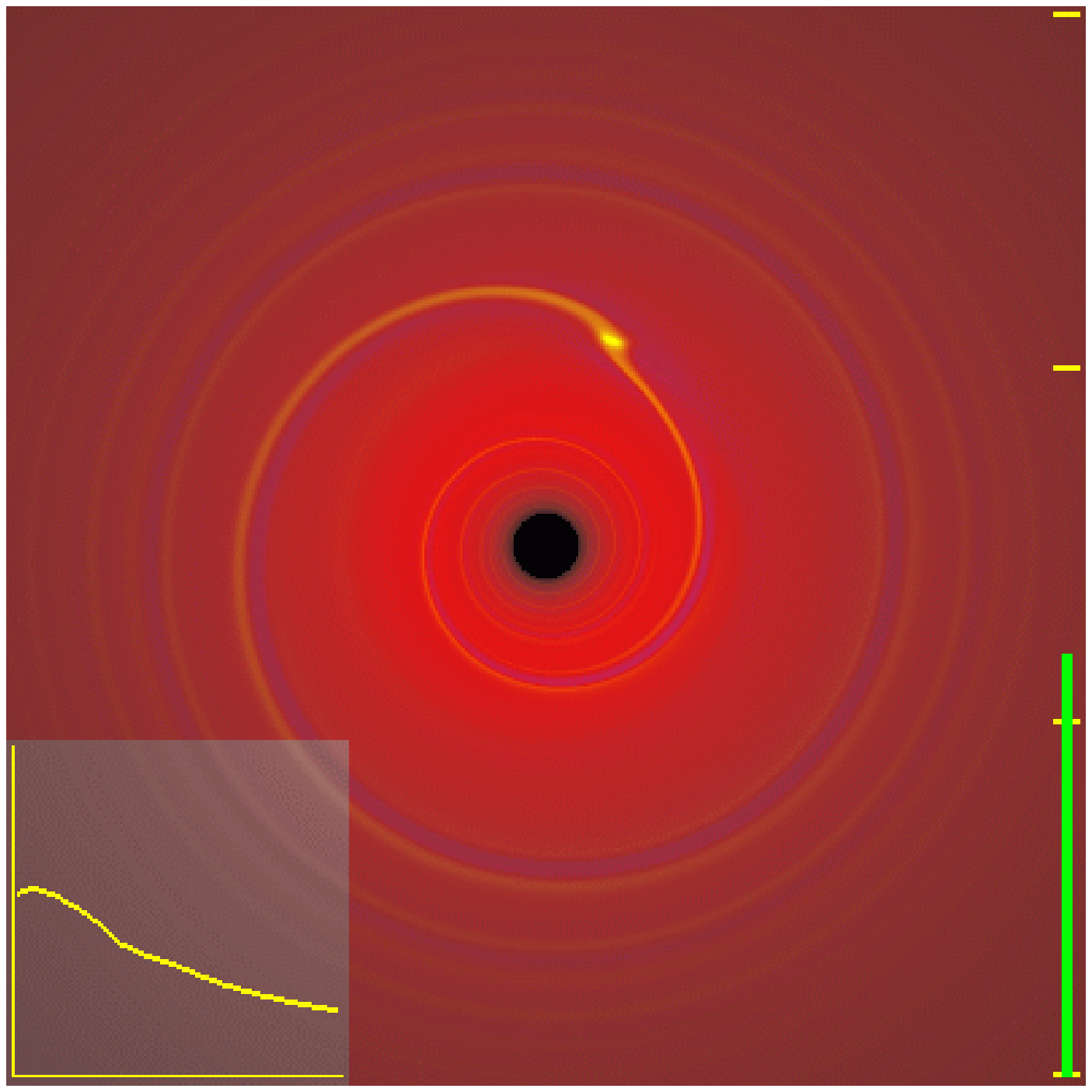}
\includegraphics[scale=0.35]{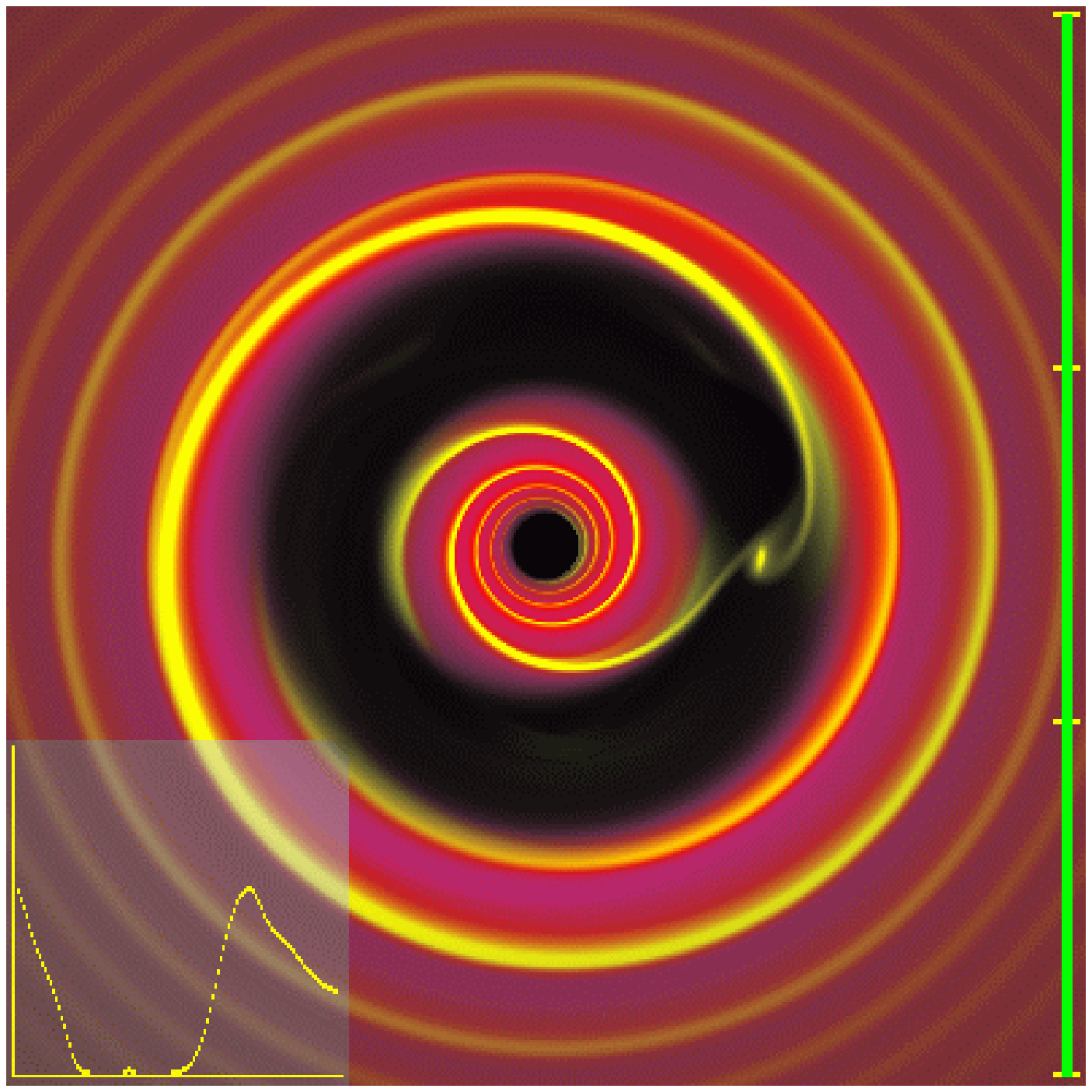}
\caption{Snapshots of the gas density in a PP disk when perturbed by
  an earth mass planet ({\it left panel}) and a Jupiter mass planet
  ({\it right panel}). The planet undergoes type I migration in the first
  case and type II migration in the second case. Inserts on the lower left
  hand side of each panels plot the azimutally averaged radial profile
  of the disk surface density. Images taken from P.Armitage website
  (\protect\url{http://jila.colorado.edu/~pja/planet_migration.html}). }
\label{planet&disk_fig}
\end{center}
\end{figure}

\paragraph{Planet/disk interaction in turbulent PP disks} At the
beginning of their lifes, young planets are still embedded in the PP
disks in which they are born. Their gravitational potential adds up to
that of the central star and perturb the disk structure. One
manifestation of that perturbation takes the form of density waves
that are excited in the disk. These waves, in turn, gravitationally
torque the planet with the result of modifying their angular momentum
and of changing their semi major axis: young protoplanets migrate
radially (usually inward) in PP 
disks. The details of the interaction between the disk and the
planet are complex and beyond the scope of this lecture. It has been
the focus of intense research in the last decades and major reviews
have been written in the last few years, to which the interested
readers is referred \citep[see, e.g.][and references
  therein]{baruteau&masset13}. For the purpose of this 
lecture, it will be enough to know that this interaction can be
divided in two broad categories that depend on planet mass:
\begin{enumerate}[topsep=0pt,partopsep=0pt,leftmargin=10pt,itemsep=-2pt]
\item[$\bullet$] Type I planet/disk interaction, relevant to low mass
  planet (typically earth to Neptune mass planets): in this case, the
  waves excited by the planet are linear in 
  amplitude. They exert a torque on the planet resulting in its inward
  migration on timescales that can be as small as $10^5$
  years. This fast migration represents a serious challenge for planet
  formation theory as the typical dissipation timescale of PP disks
  is longer by about an order of magnitude. Planet undergoing
  type I migration should then quickly reach the very vicinity of their
  central star, and perhaps even be swallowed by it (depending on the
  presence of an inner magnetospheric cavity). This problem has
  stimulated recent interest in 
  studying the interaction between the planet and the gas located in
  its coorbital region, the so--called corotation torque. In some 
  circumstances, it has been found that the corotation torque can
  reduce and even reverse type I migration.
\item[$\bullet$] Type II planet/disk interaction, relevant to massive
  planets (with masses typically in the range of that of giant gaseous
  planets): in this case, the waves excited by the planet are
  nonlinear. Shocks form and deposit their angular momentum in the
  planet vicinity. Matter is evacuated from the region coorbiting with
  the planet, creating a gap (i.e. a low density annular region)
  around the planet orbit. The planet is locked inside that gap and
  follows the disk evolution. Planet migration occurs on viscous
  timescales in this case.
\end{enumerate}
The structure of the disk in both situations is illustrated in
figure~\ref{planet&disk_fig}. For completeness, it should be
mentioned that a third regime exists, type III migration
\citep{masset&pap03}. It corresponds to an intermediate mass planet in
a more massive disk and can result in fast inward or outward
migration. We shall not considered that case any further in the
present lecture.

\noindent
Most of the numerical simulations of that problem published so far
have focused on planet/disk interaction in situations where the disk
flow is laminar and viscous. In such simulations, viscosity is
included as a large scale model of the turbulence. This approach
significantly simplifies the problem. Simple situations (2D, small
mass limit) are amenable to an analytical treatment that is very
useful to help interpret the simulations. It also greatly reduces the
computational cost associated with numerical simulations of the
problem. However, it should not be forgotten that the viscous
treatment of the dissipation in the disk is only a model of the effect
of turbulence. As such, it 
has limits and can fail in some circumstances. Fortunately,
thanks to the large increase in computational resources of the last
few years, it is now possible to start addressing the question of the
validity of that approach and to investigate the
peculiarities, if any, introduced by the fact that the flow is
turbulent and not laminar. The purpose of this section is to review
this on--going effort.

\begin{figure}
\begin{center}
\includegraphics[scale=0.75]{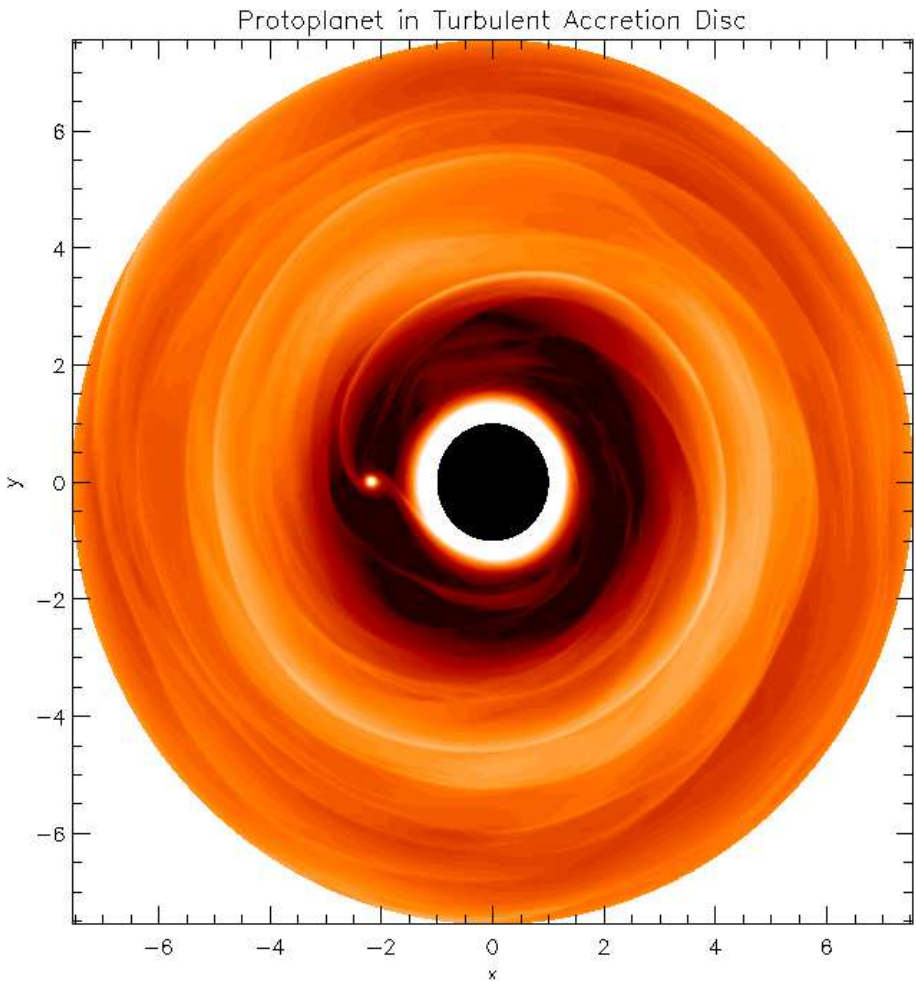}
\caption{Midplane gas density in a turbulent PP disks perturbed by a
  Jupiter mass planet. Note the presence of turbulent density
  fluctuations that perturbs the propagation of density waves excited
  by the planet. Despite the presence of the turbulence, the disk
  structure remains similar to that of a laminar disk such as shown on
  figure~\ref{planet&disk_fig}. From \citet{nelson&pap03b}}
\label{typeII_turb_fig}
\end{center}
\end{figure}

\noindent
The overall picture that emerges is as follows: in general, the results
obtained using a viscous model are reasonable. They always
produce results that are in good qualitative agreement with the simple
2D viscous approach: for example, the disk structure as perturbed by a
Jupiter mass planet in a turbulent disk is illustrated in
figure~\ref{typeII_turb_fig}. Aside from the turbulent density
fluctuations, there is good agreement with the laminar 
case (right panel of figure~\ref{planet&disk_fig}). The quantitative
agreement between the two types of approaches is also acceptable in
most situations. For example, the migration rate of a 30 earth mass
planet in a turbulent disk agrees well with the expectations based on
viscous and laminar disk simulations \citep{nelson&pap04b}. The gap
structure is also found to be well predicted by such simple
simulations \citep{nelson&pap03b,wintersetal03}\footnote{Note however
that very recent results suggest that this might not be true for all
magnetic field topologies \citep{zhuetal13}.}

\begin{figure}
\begin{center}
\includegraphics[scale=0.51]{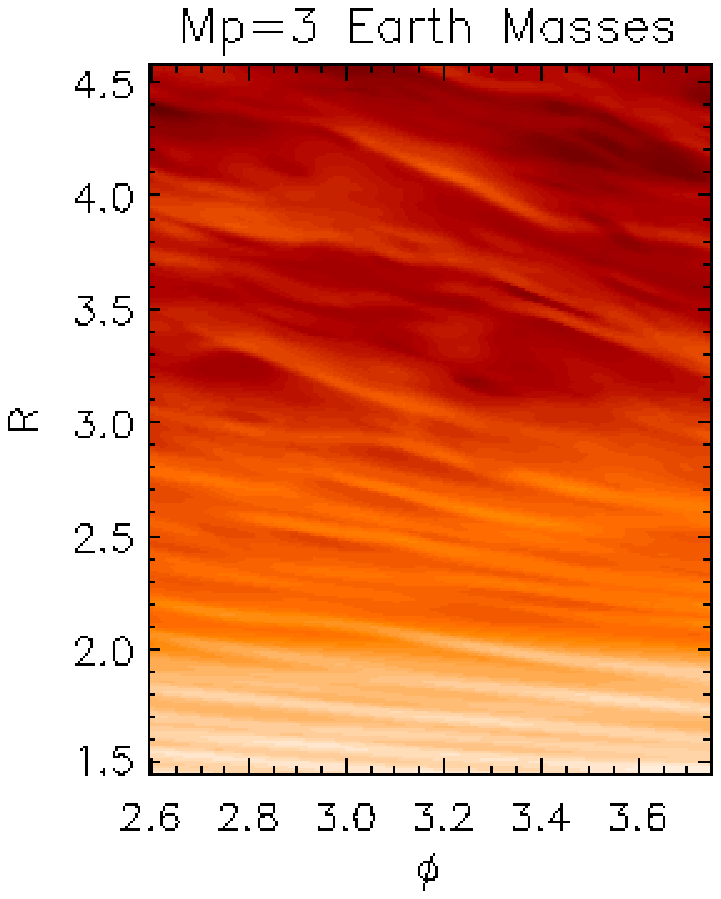}
\includegraphics[scale=0.51]{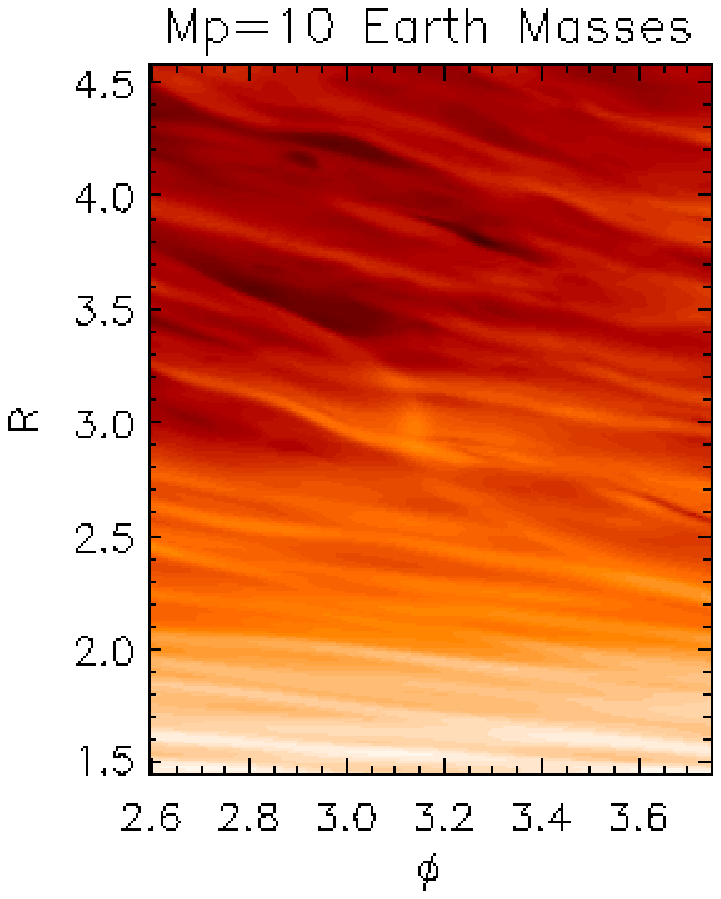}
\includegraphics[scale=0.51]{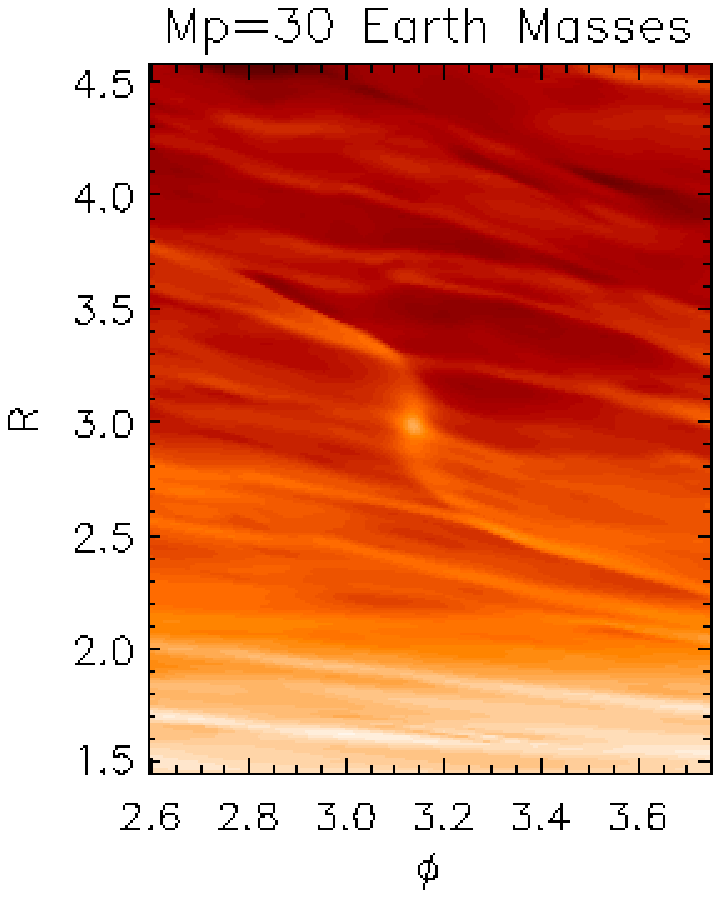}
\caption{Close--up view of the gas density in a PP disk perturbed by a
  planet of mass $M_p=3$ M$_{\oplus}$ ({\it left panel}), $M_p=10$
  M$_{\oplus}$ ({\it middle panel}) and $M_p=30$ M$_{\oplus}$ ({\it
    right panel}). Note the change in the relative importance of the
  planetary wake and of the turbulent density fluctuations as the
  planet mass is varied. In the first case, planet migration is
  completely dominated by the stochastic torque caused by these
  turbulent fluctuations. From \citet{nelson&pap04b}} 
\label{typeI_turb_fig}
\end{center}
\end{figure}

However, there are also differences some of which are important in the
context of planet formation theory: 
\begin{enumerate}[topsep=0pt,partopsep=0pt,leftmargin=10pt,itemsep=-2pt]
\item[$\bullet$] When the planet mass is small, the amplitude of the
  density wave it excites is of the same order as that of the
  turbulent density fluctuations (see for example
  figure~\ref{typeI_turb_fig}). This means that the torque of the 
  later, which is random, is of comparable amplitude as that of the
  former. Because it is fluctuating in time, the total torque exerted
  by the disk on the planet is then the sum of a constant component and a
  fluctuating component. As a result the planet experiences a random
  walk in addition to its systematic inward migration. This process is
  now called stochastic migration \citep{nelson&pap04b}. It offers a
  potential solution to the inward migration problem of small mass
  planets by reducing type I migration rates, even if the
  extrapolation of the simulation results to long evolutionary
  timescales is still debated.
\item[$\bullet$] The corotation torque arises from scales comparable to
  the disk scaleheight, or, equivalently, comparable to the scale of
  the turbulence. This suggests that a diffusive description of the
  effect of the turbulence is questionable. Recent global disk
  simulations of planet/disk interaction in turbulent disk aiming at
  isolating the effect of the corotation torque have been able to 
  show that the coorbital region still creates a torque onto the
  planet despite the presence of the turbulence
  \citep{baruteauetal11}. Nevertheless, the properties of that 
  torque (i.e. its scaling with the disk physical parameters and with
  the magnetic field) display departure from the predictions of a pure
  viscous model that should be investigated in future simulations
  systematically covering a wider range of conditions.
\end{enumerate}
It should be emphasize that published studies of the effect of MHD 
turbulence on planet/disk interaction have so far been limited to
idealized situations. A major challenge of future years will be to
investigate the modifications of this picture that result from the
complex structure of PP disks highlighted in previous sections.

\section{Conclusions}

At the time of concluding this lecture, it is important to emphasize
once more that it has no ambition but being a concise (and thus,
incomplete) introduction to the question of MRI--driven angular
momentum transport. Several aspects of the problem
are only partially covered or even completely ignored. This is
because the field is so vast. The study of angular momentum transport
in PP disks is interdisciplinary in nature. It brings together
various aspects of modern astrophysics such as plasma physics, fluid
dynamics, chemistry and radiative transfer and intimately mixes
analytical and numerical approaches. My hope is that the variety of
topics introduced in this lecture will stimulate the interest of the
reader. The references that are given along the way are as many
starting points for further readings. 

\noindent
Before closing, and maybe to stimulate discussions, controversy,
future research and hopefully progress, this lecture should end with 
an humble note. Despite the impressive achievements of the past
twenty years, it is fair to say that we still don't know with enough
confidence at which rate angular momentum is being transported in PP
disks. Their structure is still highly uncertain. As 
highlighted by the last part of this lecture, this is not without
consequences for several aspects of planet formation. Much remains to
be done before we can form a self--consistent picture of how planets 
form in the universe. 


\subsection*{Acknowledgment}
I am indebted to all the colleagues that contributed to my
understanding of the field over the past ten years. This lecture would 
never have come to be without their support and communicative 
interest for the subjects of accretion disk dynamics and planet
formation. I also acknowledge Xuening Bai, Steven Balbus, Clement
Baruteau, Arnaud Belloche and Geoffroy Lesur for a careful reading of 
an earlier draft of this lecture.

{\small
\bibliographystyle{astron}
\bibliography{author}
}

\end{document}